\documentclass[a4paper,twocolumn,prd,superscriptaddress,showpacs,floatfix%
,preprintnumbers,nofootinbib]{revtex4}

\usepackage{verbatim}
\usepackage{epsfig}
\usepackage{bm}
\usepackage[usenames]{color}

\newcommand{\D}{{\rm d}}
\newcommand{\I}{{\rm i}}
\newcommand{\E}{{\rm e}}

\begin{document}

\title{Fast time variations
of supernova neutrino fluxes and their detectability}

\author{Tina Lund}
\affiliation{Department of Physics and Astronomy,
Aarhus University, Ny Munkegade 120, 8000 Aarhus C, Denmark}

\author{Andreas Marek}
\affiliation{Max-Planck-Institut f\"ur Astrophysik,
Karl-Schwarzschild-Str.~1, 85748 Garching, Germany}

\author{Cecilia Lunardini}
\affiliation{Arizona State University, Tempe, AZ 85287-1504, USA}
\affiliation{RIKEN BNL Research Center, Brookhaven National
  Laboratory, Upton, NY 11973, USA}

\author{Hans-Thomas Janka}
\affiliation{Max-Planck-Institut f\"ur Astrophysik,
Karl-Schwarzschild-Str.~1, 85748 Garching, Germany}

\author{Georg Raffelt}
\affiliation{Max-Planck-Institut f\"ur Physik
(Werner-Heisenberg-Institut), F\"ohringer Ring 6, 80805 M\"unchen,
Germany}

\date{18 August 2010}

\preprint{MPP-2010-47}

\begin{abstract}
In the delayed explosion scenario of core-collapse supernovae (SNe),
the accretion phase shows pronounced convective overturns and a
low-multipole hydrodynamic instability, the standing accretion shock
instability (SASI). These effects imprint detectable fast time
variations on the emerging neutrino flux. Among existing detectors,
IceCube is best suited to this task, providing an event rate of
$\sim1000~{\rm ms}^{-1}$ during the accretion phase for a fiducial
SN distance of 10~kpc, comparable to what could be achieved with a
megaton water Cherenkov detector. If the SASI activity lasts for
several hundred~ms, a Fourier component with an amplitude of 1\% of
the average signal clearly sticks out from the shot noise. We
analyze in detail the output of axially symmetric hydrodynamical
simulations that predict much larger amplitudes up to frequencies of
a few hundred Hz. If these models are roughly representative for
realistic SNe, fast time variations of the neutrino signal are easily
detectable in IceCube or future megaton-class instruments. We
also discuss the information that could be deduced from such a
measurement about the physics in the SN core and the explosion
mechanism of the SN.
\end{abstract}


\maketitle

\section{Introduction}

The delayed explosion scenario remains the standard paradigm for the
core collapse supernova (SN) mechanism. After core bounce a shock
wave forms that stalls at a typical radius of 100--200~km while
matter keeps falling in, forming a standing accretion shock that can
last for several hundred ms before the shock is re-launched,
presumably after sufficient neutrino energy deposition in the region
behind the shock. Two- and three-dimensional hydrodynamic
simulations reveal convective instabilities that quickly develop
into large-scale convective overturns and a strong dipole
oscillation of the neutron star against the ``cavity'' formed by the
standing shock, the standing accretion shock instability
(SASI)~\cite{Blondin:2002sm, Ohnishi:2005cv, Foglizzo:2007,
Scheck:2008, Marek:2007gr, Marek:2008qi, Ott:2008jb}.

\begin{figure}[t!]
\resizebox{80mm}{!}{\includegraphics{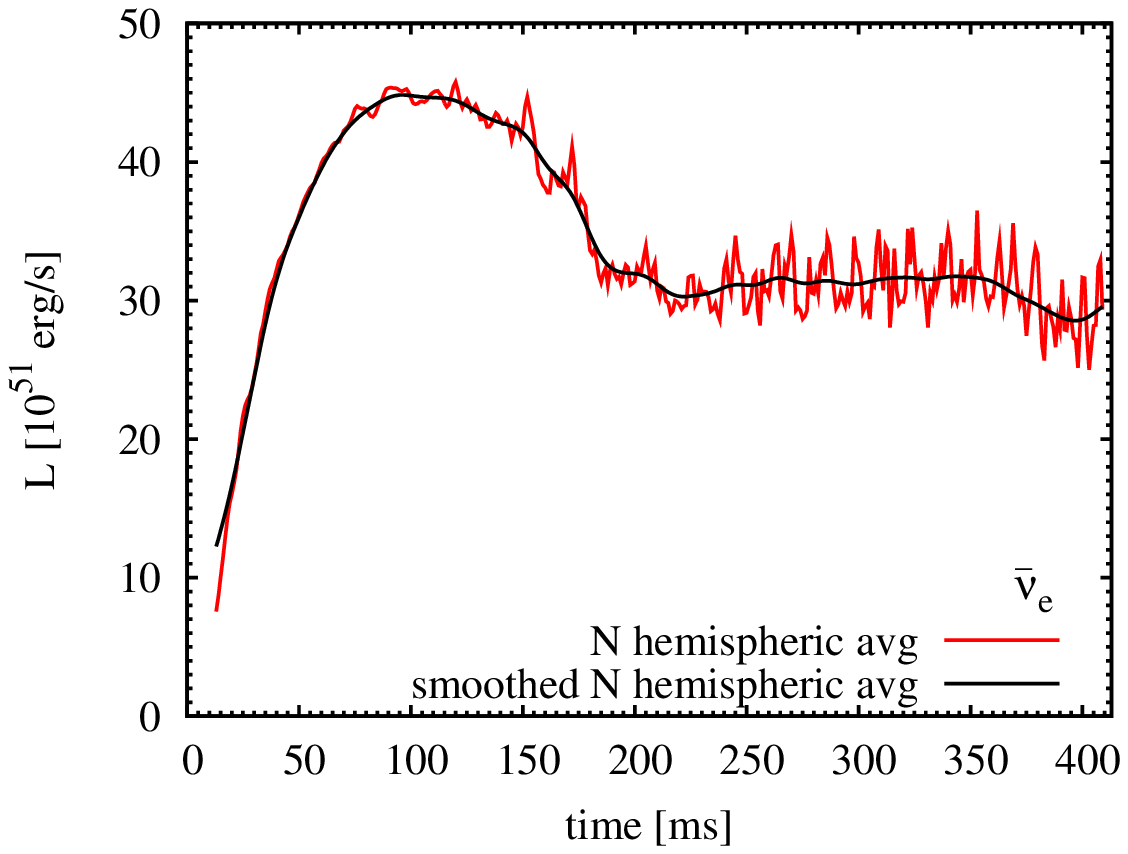}}\\
\resizebox{80mm}{!}{\includegraphics{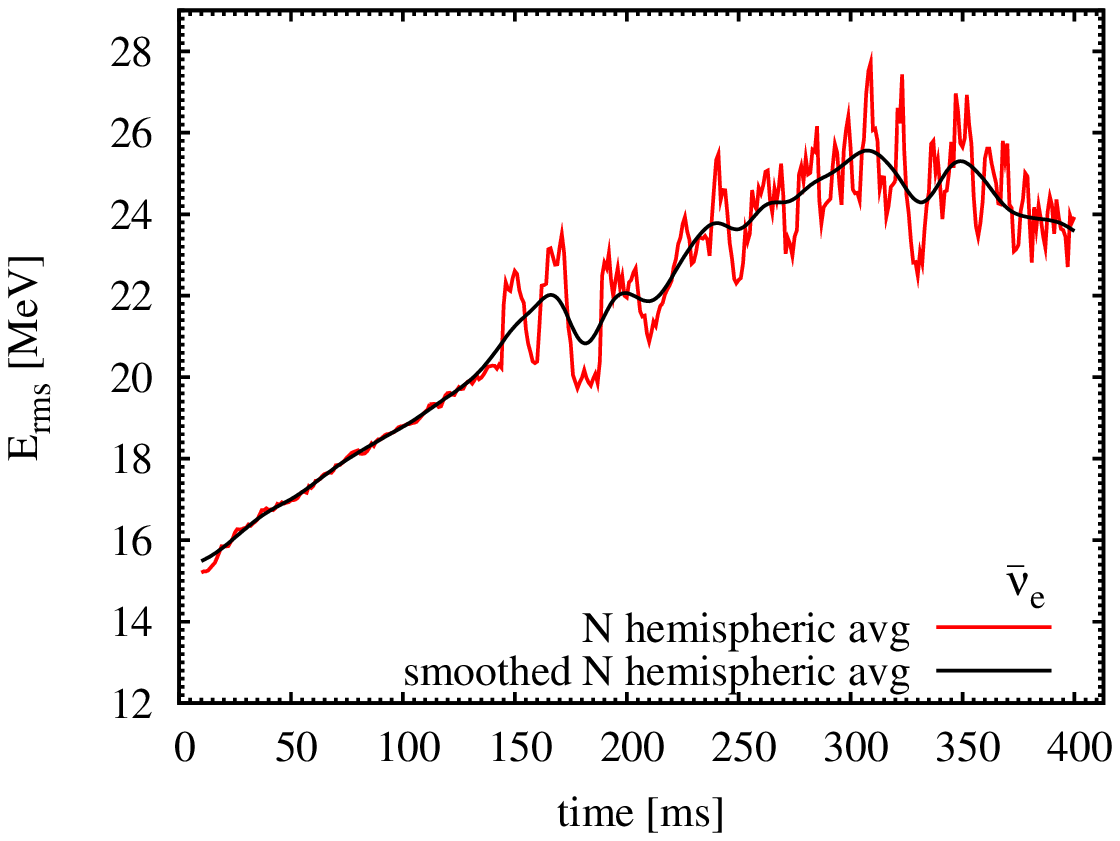}}\\
\resizebox{80mm}{!}{\includegraphics{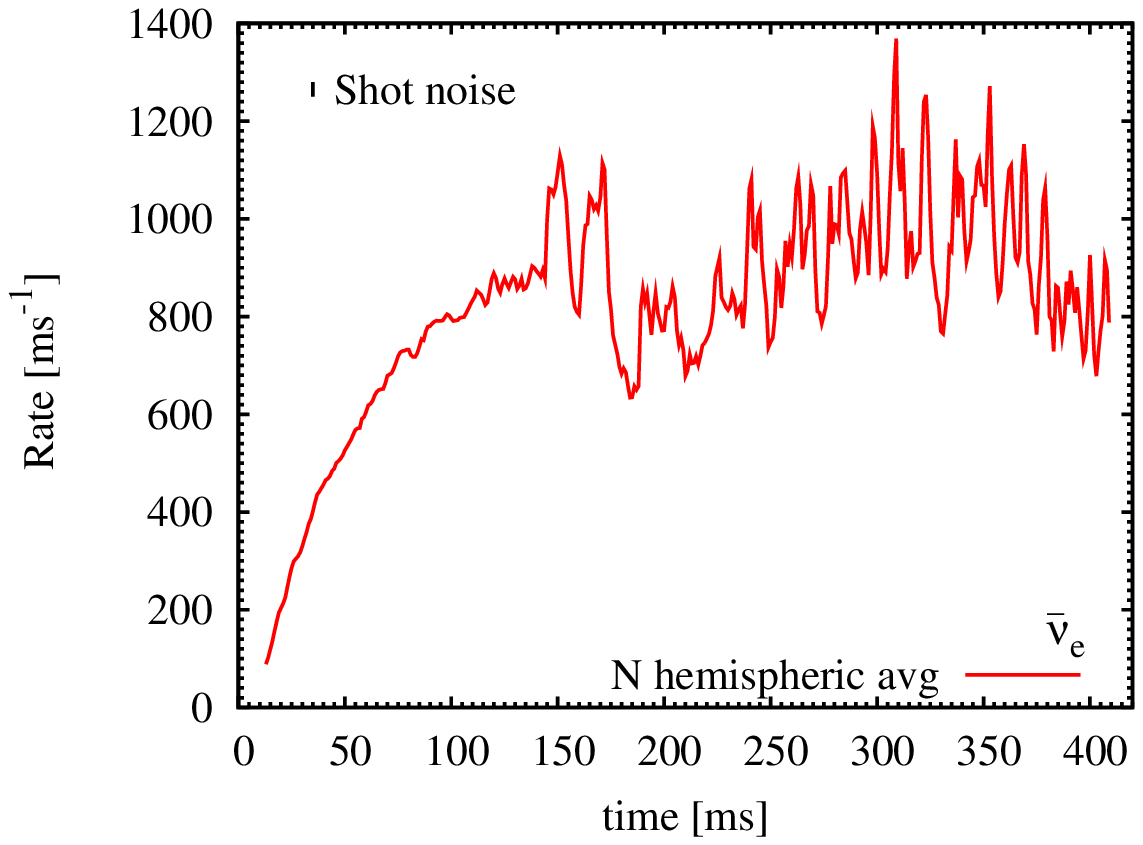}}
\caption{\label{fig:firstexample} {\it Top:} $\bar\nu_e$ luminosity
of our baseline SN model sampled at 1~ms intervals.
Red line: North hemispheric average.
Black line: Moving average with a Gaussian window function
($\sigma = 7$~ms).
{\it Middle:} $\bar\nu_e$ rms energy. Red and black lines as the panel above.
{\it Bottom:} Detection rate in IceCube. Also shown is
the $1\,\sigma$ range caused by shot noise, assuming a bin width of 1~ms.}
\vskip-3pt
\end{figure}

During the accretion phase, neutrino emission is particularly large,
being powered primarily by the gravitational energy of the
in-falling material. In the SASI scenario the neutrino emission is
strongly modulated. As an example we show in
Fig.~\ref{fig:firstexample} the $\bar\nu_e$ luminosity as a function
of time, averaged over one hemisphere, from the two-dimensional
simulations of Marek, Janka and M\"uller (2009) \cite{Marek:2008qi}.
As noted by these authors, such large flux variations could well
become detectable in the high-statistics neutrino signal of the next
galactic SN, revealing direct evidence for the predicted SASI mode
and convective overturns.

\begin{figure}[t!]
\resizebox{80mm}{!}{\includegraphics{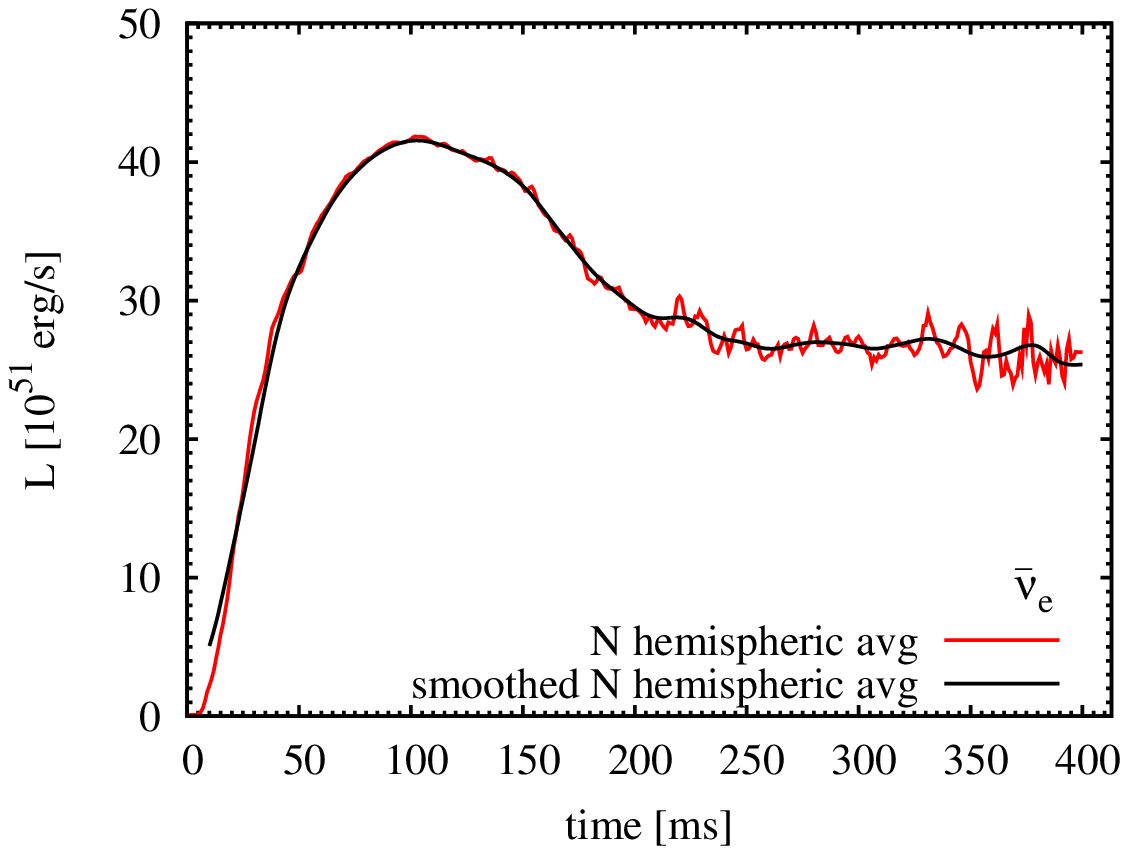}}\\
\resizebox{80mm}{!}{\includegraphics{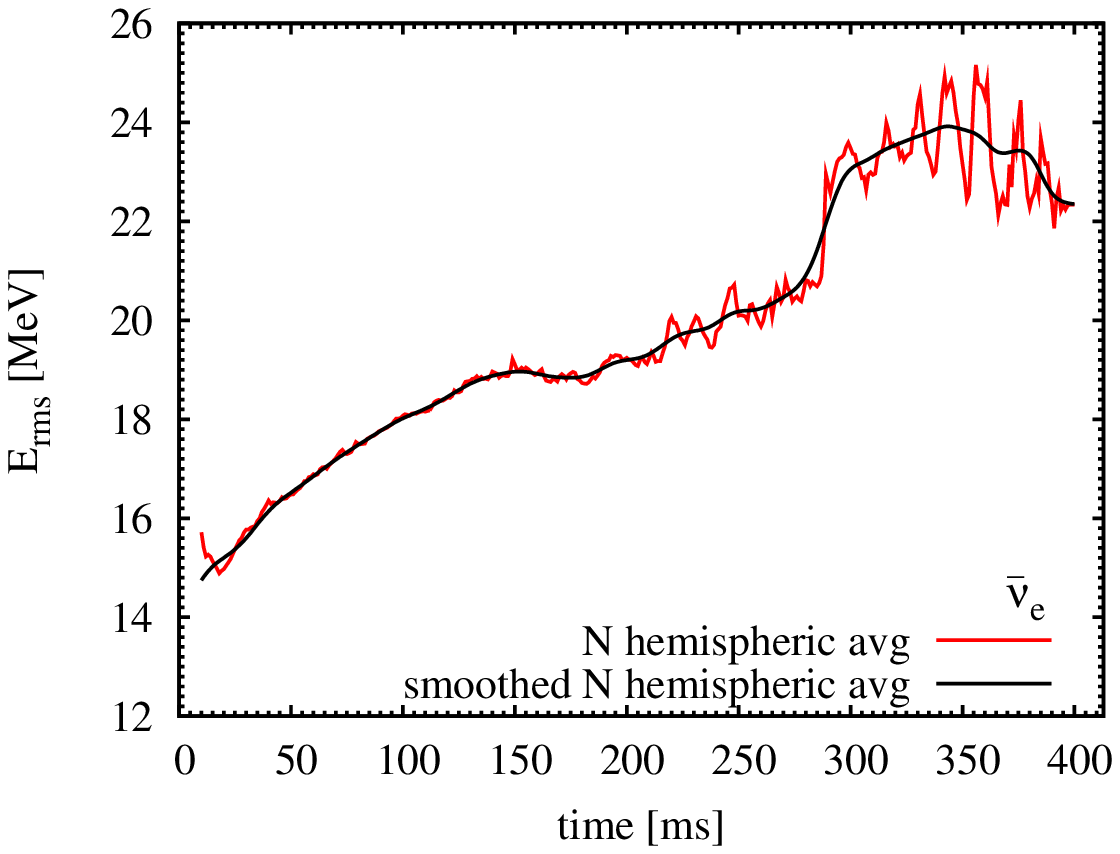}}\\
\resizebox{80mm}{!}{\includegraphics{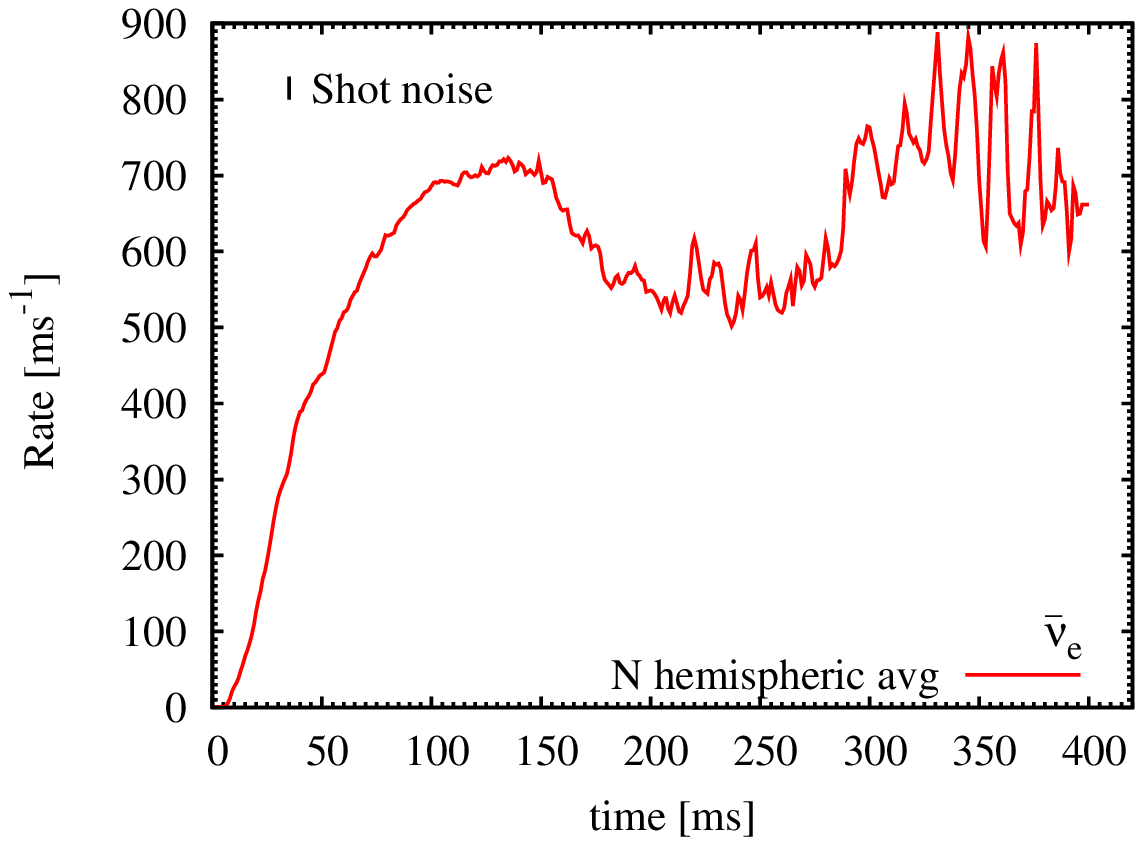}}
\caption{\label{fig:firstexampleX} Same as Fig.~\ref{fig:firstexample}
using the EoS of Hillebrandt and Wolff.}
\vskip-3pt
\end{figure}

The detection of fast time variations, or equivalently, identifying
high-frequency Fourier modes in the signal, is limited by the number
of registered events: A significant signal must stick above the shot
noise caused by the fluctuating event rate, so a large counting rate
is crucial. Among the existing or near-future detectors, IceCube is
the most promising because it detects a large number of Cherenkov
photons triggered by neutrinos. At most one single Cherenkov photon
is picked up from a given neutrino, so every photon tags the arrival
time of a different neutrino. For our SN example of Fig.~1, assumed
at a fiducial distance of 10~kpc, the maximum photon detection rate
is roughly $1000~{\rm ms}^{-1}$, similar to the intrinsic background
rate. On the other hand, for Super-Kamiokande (fiducial volume
22.5~kt) the corresponding neutrino detection rate is approximately
two orders of magnitude smaller, although essentially background
free. Based on this simple estimate, we use IceCube as our benchmark
detector. On the other hand, a megaton-class water Cherenkov
detector would achieve neutrino detection rates similar to IceCube
and in addition would provide event-by-event energy information, a
quantity also showing strong fluctuations. The potential of such a
detector will be studied elsewhere.

A possible limiting factor to detecting fast signal variations is
time-of-flight dispersion caused by neutrino masses. The delay of
arrival times is
\begin{equation}
\Delta t=0.57~{\rm ms}\,\left(\frac{m}{\rm eV}\right)^2\,
\left(\frac{30~{\rm MeV}}{E}\right)^2\,
\left(\frac{D}{10~{\rm kpc}}\right)^2\,.
\end{equation}
We will see that signal variations may be detectable up to a few
hundred Hz, corresponding to time scales of several ms. So even for
eV neutrino masses, arrival time dispersion would be a marginal
effect. Moreover, current cosmological limits on the overall
neutrino mass scale are approximately 0.2~eV
\cite{Hannestad:2010yi}, very similar to the sensitivity of the
ongoing KATRIN experiment~\cite{Drexlin:2008zz}. Assuming KATRIN
will confirm this limit, arrival time dispersion of SN neutrinos
will be completely irrelevant in our context. Should KATRIN discover
eV-scale neutrino masses in violation of cosmological limits one
could return to this study and include time-of-flight dispersion.

In Sec.~\ref{sec:detector} we review the detector response of
IceCube to SN neutrinos. In Sec.~\ref{sec:SNsignal} we use the
output from the numerical models of Marek, Janka and M\"uller (2009)
\cite{Marek:2008qi} and study the signal power as a function of
frequency relative to the detector shot noise. In
Sec.~\ref{sec:correlations} we briefly discuss the fluctuations of
neutrino energies relative to luminosity variations. In
Sec.~\ref{sec:HWeos} we consider the modification caused by a
stiffer nuclear equation of state. In Sect.~\ref{sec:implications}
we interpret the results of our analysis on the basis of present
explosion models and our understanding of the physics relevant in
the SN core. We also briefly address the question what could be
learned if SASI and convective neutrino signal variations were
detected. We discuss and summarize our findings in
Sec.~\ref{sec:conclusions}. In Appendix~\ref{sec:shotnoise} we
derive the detector shot noise and define the normalization of our
Fourier transform of binned~data.

\section{Detector Model}                          \label{sec:detector}

The IceCube detector, soon to be completed at the South Pole,
currently provides by far the largest detection rate for the next
nearby SN. For our sensitivity forecast we use a somewhat schematic
model for its response to a SN neutrino signal. For the finished
detector with 4800 optical modules, the latest efficiencies provide
a detection rate of Cherenkov photons originating from the dominant
inverse beta reaction $\bar\nu_e+p\to n+e^+$
of~\cite{Dighe:2003b,Kowarik:2009qr, Halzen:2009sm}
\begin{equation}\label{eq:eventrate}
R_{\bar\nu_e}=114~{\rm ms}^{-1}\,
\frac{L_{\bar\nu_e}}{10^{52}~{\rm erg}~{\rm s}^{-1}}
\left(\frac{10~{\rm kpc}}{D}\right)^2
\left(\frac{E_{\rm rms}}{15~{\rm MeV}}\right)^2\,.
\end{equation}
Here we use the definition
\begin{equation}
E_{\rm rms}^2=\frac{\langle E^3\rangle}{\langle E\rangle}\,,
\end{equation}
where the average is to be taken over the neutrino distribution
function. This quantity plays the role of an rms energy relative to
the energy spectrum, not the flux spectrum. Other authors use the
definition $E_{\rm rms}=\sqrt{\langle E^2\rangle}$, which is perhaps 
more appropriately called the rms energy, but our definition is what
appears in the IceCube rate and thus will be used. In other words,
because the Cherenkov light measures the neutrino energy deposition
in ice, $L_{\bar\nu_e}$ and $E_{\rm rms}$ are the most natural
parameters to describe the instantaneous neutrino flux. Our estimate
of the photon count rate Eq.~(\ref{eq:eventrate}) uses an
approximate inverse beta cross section of
$\sigma=9.52\times10^{-44}~ {\rm cm}^2\,(E_{\bar\nu_e}/{\rm MeV})^2$
to obtain a simple scaling behavior with energy.

For low-energy neutrino detection, IceCube is a very coarse
detector, implying that from a given neutrino it picks up at most
one Cherenkov photon. Assuming more closely spaced optical modules,
the average photon detection rate remains unchanged, but the
fluctuations increase in that several detected photons may have been
triggered by the same neutrino. This increased shot noise reduces
the capability to detect fast time variations. In the extreme case
of a densely instrumented detector such as Super-Kamiokande, one
measures so many photons from a given neutrino that one can
reconstruct detailed energy and directional information, at the
expense of relatively few neutrino events and therefore much larger
shot noise. In this sense the apparent weakness of IceCube for
low-energy neutrino detection is actually a virtue for diagnosing
fast time variations. Since for IceCube a single photon detection is
identical with detecting the arrival time of a neutrino (except for
background), we use the term ``event'' interchangeably for ``photon
detection'' or ``neutrino detection.''

For our SN models, a typical rate during the accretion phase is
around $10^3~{\rm ms}^{-1}$. This is to be compared with the
estimated IceCube background rate of~\cite{Kowarik:2009qr}
\begin{equation}\label{eq:backgroundrate}
R_{\rm bkgd}=1.34\times10^{3}~{\rm ms}^{-1}\,.
\end{equation}
This is the dark current of 280~s$^{-1}$ per optical module,
multiplied with 4800 optical modules of the final detector to be
completed by the end of this year. Therefore, for a SN at the
fiducial distance, the signal and background rates are comparable,
but the background dominates. Therefore, it is essentially the shot
noise of the background that limits the detectability of fast signal
variations unless the SN is closer.

IceCube samples the data in 1.6384~ms bins whereas low-energy water
Cherenkov or scintillator detectors register the times $t_j$ of
every event with high precision. The bin width, or the absence of
binning, only affects the sensitivity to frequencies that we will
see are too high to be detected. Therefore, the details of signal
binning do not enter our discussion. We have found it convenient to
use 1~ms as a nominal bin width and also as a sampling rate of the
numerical SN results.

The main obstacle to detecting fast time variations is shot noise
(Poisson fluctuations of the limited number of events). In the lower
panel of Fig.~\ref{fig:firstexample} we show as a vertical bar the
$1\,\sigma$ fluctuation per 1~ms bin for an event rate of
900~ms$^{-1}$. Therefore, it is evident that with IceCube one can
follow the overall neutrino light curve with excellent precision.

To estimate the required amplitude for a fast periodic variation to be
detectable, we model the signal as a sequence of arrival times $t_j$
with $j=1,\ldots,N$. The Fourier transform of this signal
(frequency $f$) is
\begin{equation}\label{eq:Fourierdefinition}
g(f)=\sum_{j=1}^{N}\E^{-\I 2\pi f t_j}
\end{equation}
with the spectral power $G(f)=|g(f)|^2$. The detection rate has
units of inverse time, so the Fourier components are dimensionless.
If the sequence of events is completely random, i.e.\ the times
$t_j$ sample a uniform distribution on a given time interval, one
can show (Appendix~\ref{sec:shotnoise}) that
\begin{equation}\label{eq:1overN}
\frac{\langle G_{f\not=0}\rangle}{G_{f=0}}=\frac{1}{N}
\end{equation}
and we note that $G(0)=N^2$.

Next we assume the signal has a frequency $f_a$ imprinted upon it,
i.e.\ it is proportional to $[1+a\cos(2\pi f_a t)]$. The power of
this signal vanishes everywhere except at $f=0$ and $f=\pm f_a$. The
relative power is $G(\pm f_a)/G(0)=a^2/4$. Therefore, an imprinted
cosine variation with the amplitude $a=2/\sqrt{N}$ is equal to the
shot noise. Assuming that the accretion phase lasts for 400~ms and
using the background rate of Eq.~(\ref{eq:backgroundrate}) as the
dominant source of shot noise, the number of events is
$N=5.4\times10^5$ and the shot-noise level corresponds to a cosine
amplitude of $a=3\times10^{-3}$. Of course, to stick reliably above
background, a Fourier component would need to be somewhat larger. In
other words, for a SN at 10~kpc one can expect to detect signal
variations with an amplitude roughly on the 1\%~level of the average
rate. The signal variations shown in the lower panel of
Fig.~\ref{fig:firstexample} would easily show up in IceCube.

\section{Numerical Supernova Signal}       \label{sec:SNsignal}

\subsection{Description of the model}

To make this rough estimate more concrete, we next use a numerical
simulation to compare the expected signal fluctuations with the
sensitivity of IceCube. The two-dimensional (axially symmetric)
simulations which this discussion is based on were performed with
the Prometheus-Vertex Code~\cite{Rampp:2002, Buras:2006a} and the
simulations were already discussed in detail in
Ref.~\cite{Marek:2008qi}. We therefore repeat only a few essential
aspects of both the numerical treatment and the simulation runs and
refer to Refs.~\cite{Marek:2008qi, Marek:2007gr} for more complete
information.

The hydrodynamic part of the code is based on a conservative and
explicit Eulerian implementation of a Godunov-type scheme with
higher-order spatial and temporal accuracy. It solves the
nonrelativistic conservation equations for the stellar fluid, whose
self-gravity is described by an ``effective relativistic potential''
\cite{Marek:2006}. It provides a sufficiently accurate approximation
of general relativistic corrections~\cite{Mueller:2010}.

The neutrino transport solver, which is coupled to the hydrodynamics
module via lepton number, energy and momentum source terms, is
computed with a ``ray-by-ray plus'' scheme \cite{Buras:2006a}. It
accounts for the full neutrino-energy dependence in the transport
but assumes the neutrino flux at every point to be radial (i.e.~the
neutrino phase space distribution function is assumed to be axially
symmetric around the radial direction), which is numerically less
demanding and more efficient than a full multi-dimensional version
of the transport.

The simulations used here are based on the progenitor model s15s7b2
from Woosley and Weaver~\cite{Woosley:1995}, and is representative
for the collapse of stars with progenitor masses around
$15\,M_\odot$. The dense proto-neutron star matter is described by
the equation of state (EoS) of Lattimer and Swesty
\cite{Lattimer:1991}, which leads to a radius of 12 km for a cold
neutron star with a gravitational mass of $1.4\,M_\odot$. We also
consider briefly an example with the EoS of Hillebrandt and Wolff
that is considerably stiffer~\cite{Hillebrandt:1985}. Unless
otherwise noted, our discussion always refers to the Lattimer and
Swesty case as a benchmark.

The two-dimensional model was computed under the assumption of axial
symmetry and covers the region between north and south pole with 192
equally spaced angular grid points. The model was evolved in total
for about 600 ms from the onset of the collapse to a time of about
450 ms after the formation of the SN shock front and shows in
the postbounce evolution a strong SASI sloshing activity of the SN
shock front.

The oscillations of the SN shock front due to SASI activity
and convective overturn cause luminosity fluctuations by modulating
the mass accretion on the proto-neutron star: a strong shock
retraction leads to a transient increase of the gas flow towards the
neutron star and to the compression and enhanced cooling of the
matter (i.e.~enhanced neutrino emission) near the neutron star
surface~\cite{Marek:2008qi}. On the other hand, a shock expansion
has the opposite effect because it causes a deceleration of the
infall or even outward acceleration of material that is accreted
through the shock front. Thus shock expansion stretches the time
this matter stays in the gain layer and less cooling by neutrino
emission occurs.

\begin{figure}[t]
\begin{tabular}{c}
\resizebox{80mm}{!}{\includegraphics{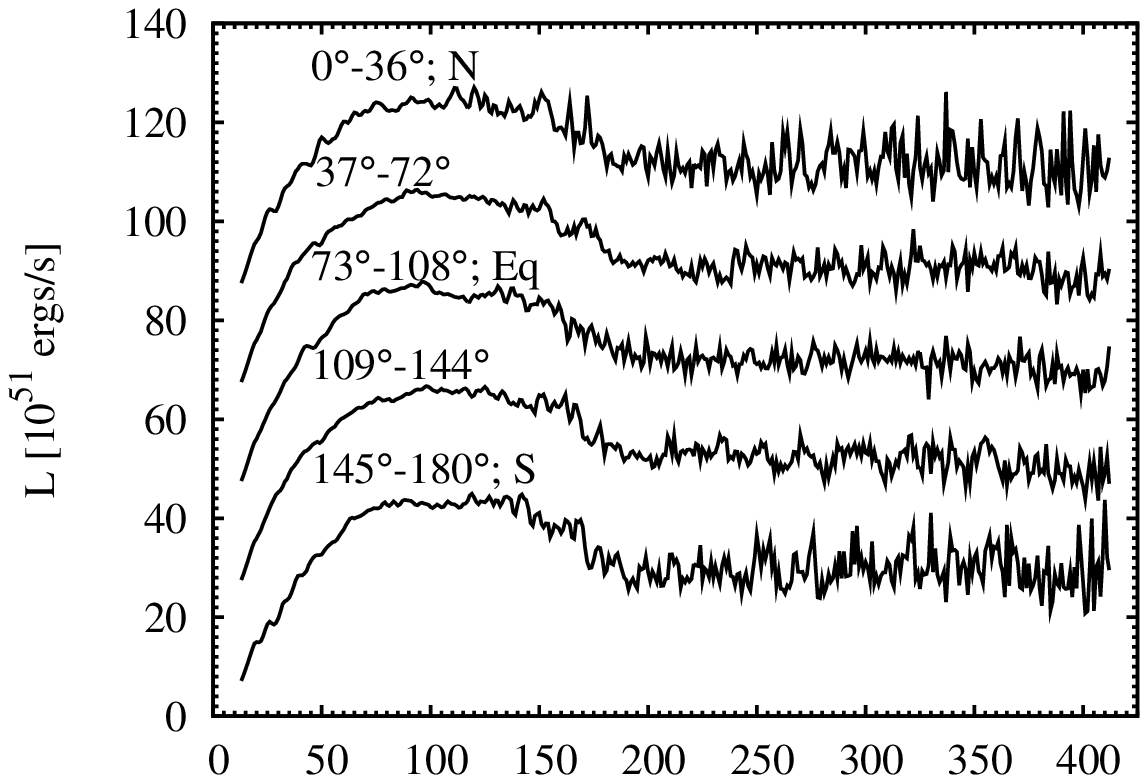}}\\
\resizebox{80mm}{!}{\includegraphics{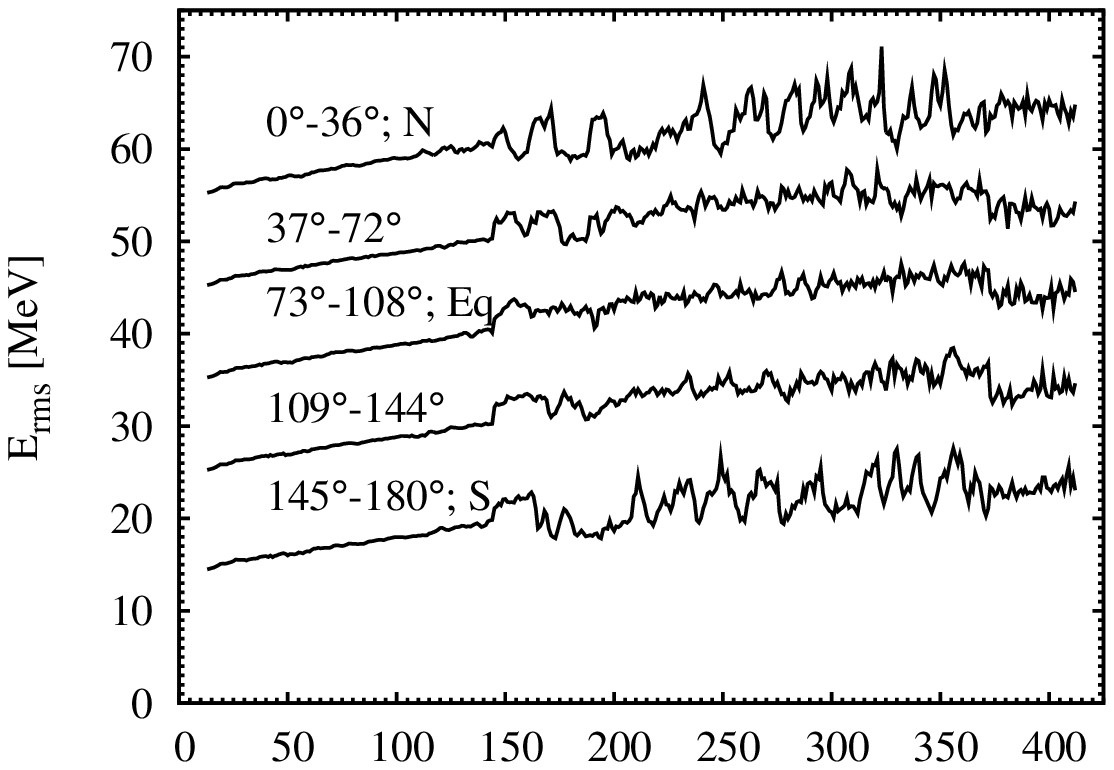}}\\
\resizebox{80mm}{!}{\includegraphics{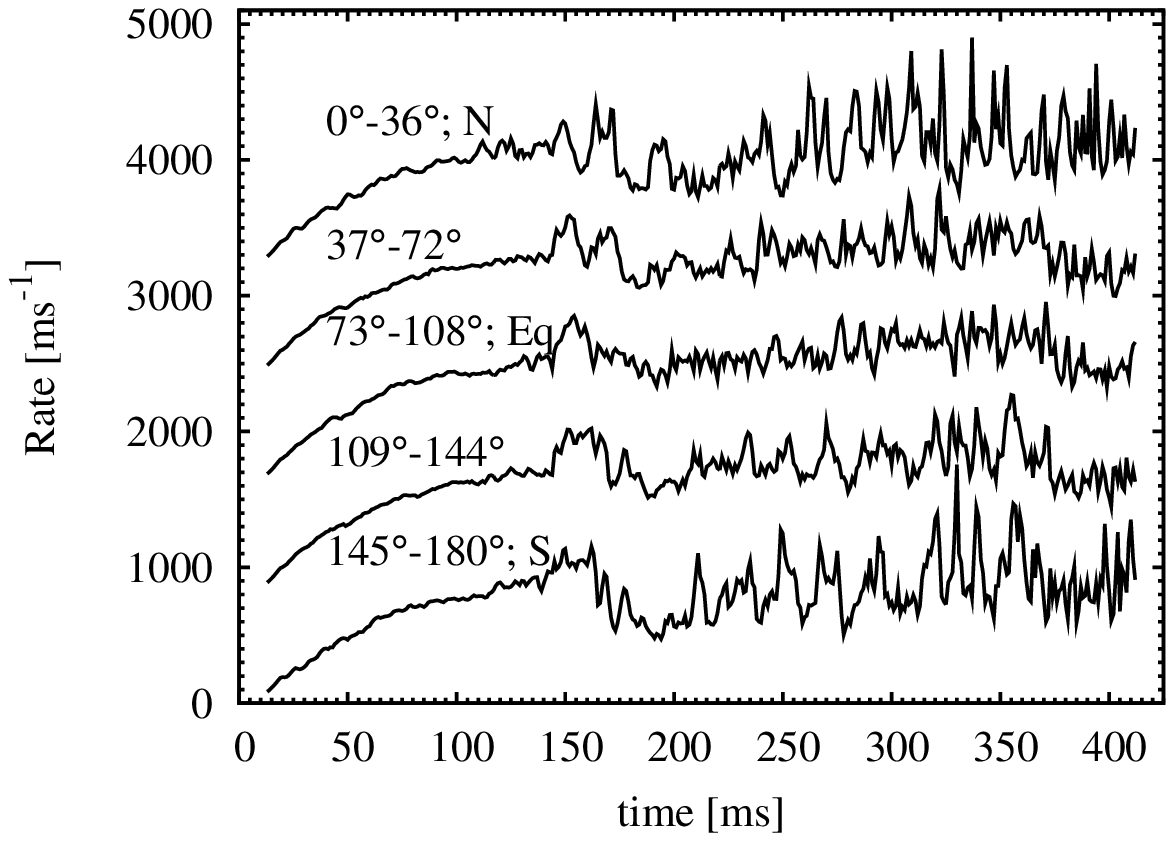}}
\end{tabular}
\caption{\label{fig:5wedge} Luminosity, rms energy and IceCube
  event rate based on the $\bar{\nu}_e$ fluxes of our model with
  Lattimer and Swesty EoS. The 96
  angular rays have been combined into five averages ranging from
  north (top curve) to south (bottom curve), where in each panel the
  curves are offset relative to each other
  by 20, 10 and 800 units of the vertical axes,
  respectively.}
\end{figure}

\begin{figure}[t]
\begin{tabular}{c}
\resizebox{80mm}{!}{\includegraphics{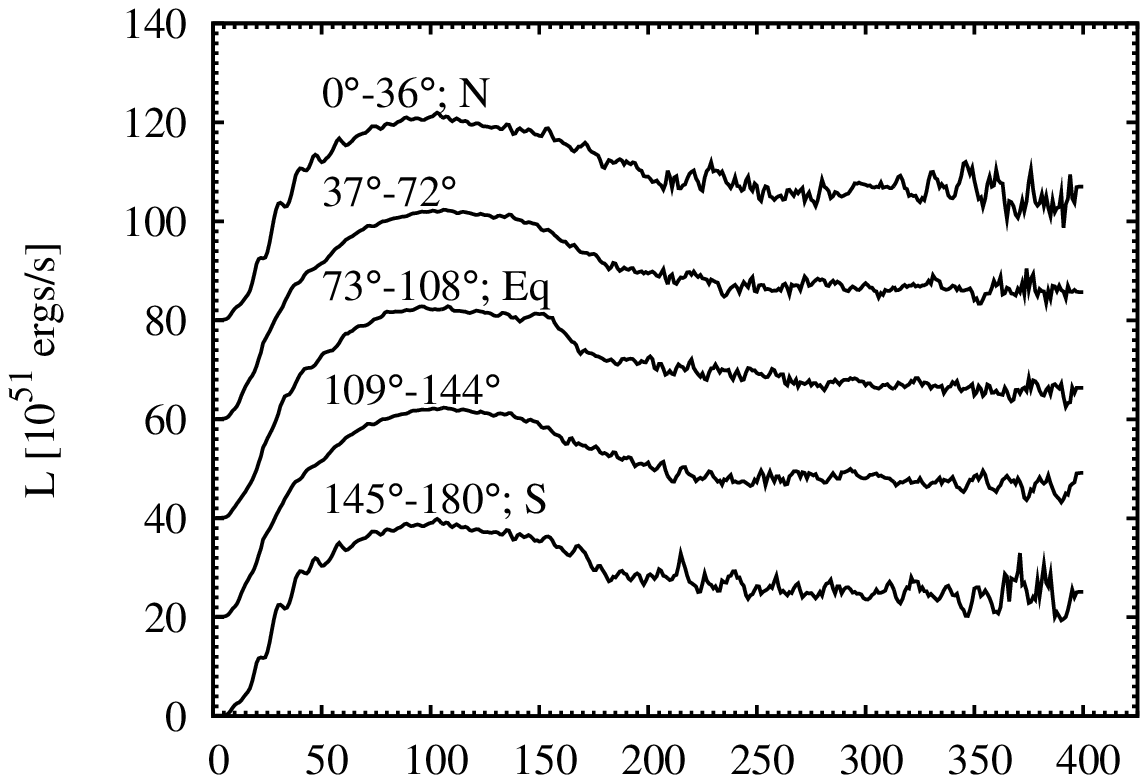}}\\
\resizebox{80mm}{!}{\includegraphics{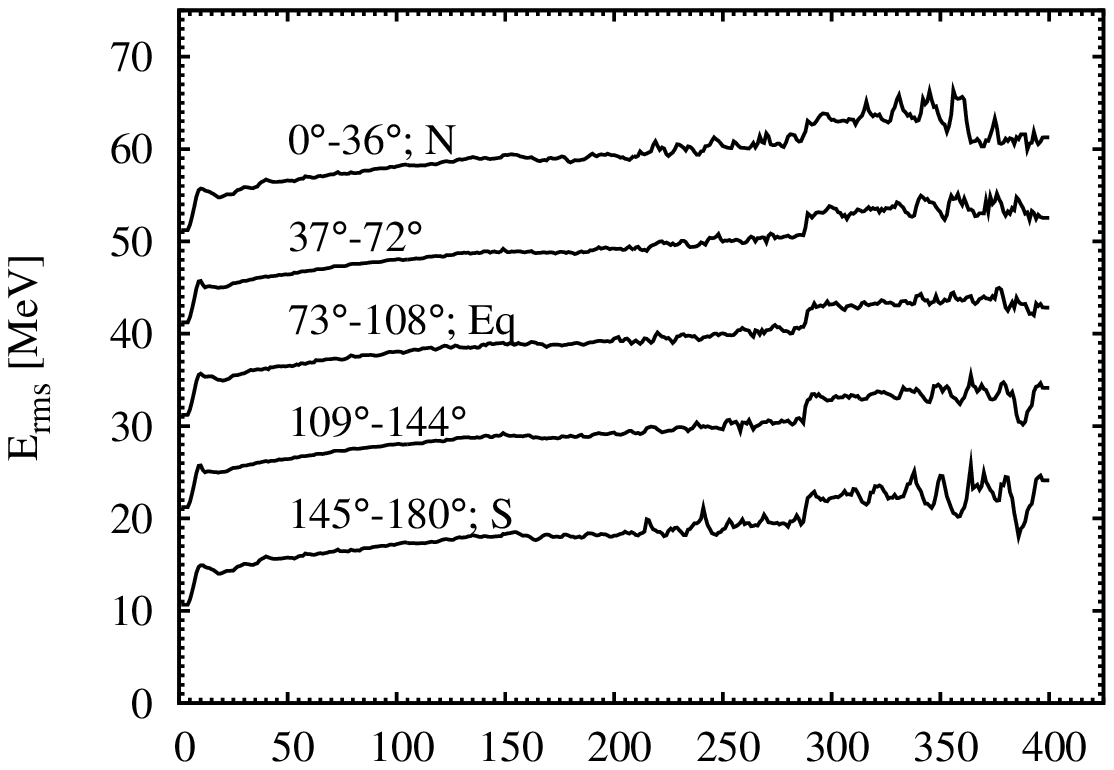}}\\
\resizebox{80mm}{!}{\includegraphics{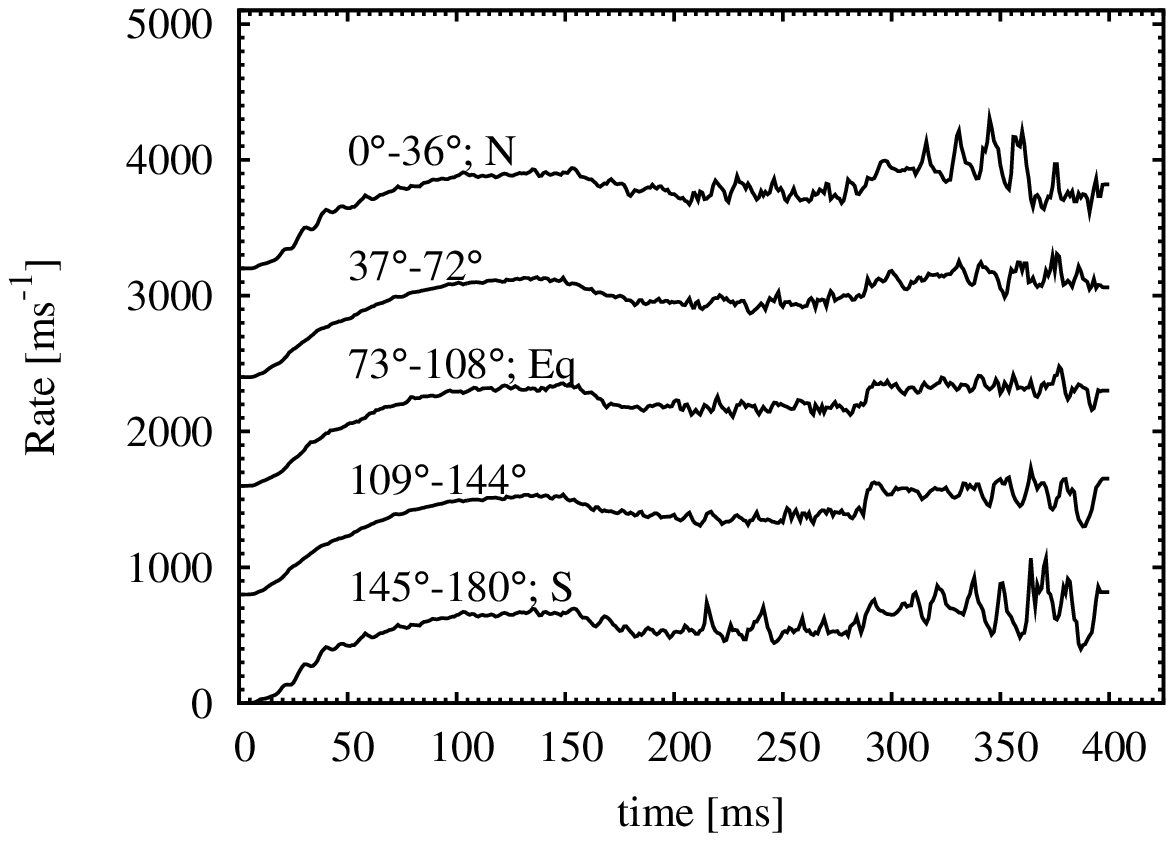}}
\end{tabular}
\caption{\label{fig:5wedgeHW} Same as Fig.~\ref{fig:5wedge} for a run
with the EoS of Hillebrandt and Wolff.}
\end{figure}

From the 192 angular rays of the models, we used in the post
processing the luminosity for all species $\nu_e$, $\bar\nu_e$ and
$\nu_x$ and the corresponding $\langle E\rangle_{\rm rms}$ on every
second angular bin and extracted the information in steps of about
0.5~ms that were subsequently resampled in exact 1~ms steps. To
illustrate the general appearance and directional differences of the
fluctuations, we combined the angular rays into 5 directional
averages: North polar, equatorial, south polar, and an intermediate
wedge between each pole and equator, each of them covering a
zenith-angle range of 36$^\circ$. For $\bar\nu_e$, the luminosity,
rms energy and IceCube detection rate of the 5 wedges are shown in
Fig.~\ref{fig:5wedge}. Of course, these plots have no direct
observational significance and merely serve to illustrate the
angular variation of the SN output. We clearly see that fluctuations
in the energy and luminosity are larger along the polar directions
than at the equator. In Fig.~\ref{fig:5wedgeHW} we show the same
information for the run with the EoS of Hillebrandt and Wolff, on
which we will comment later.

\subsection{Fourier transform and spectral power}\label{subsec:FTpower}

In order to assess the detectability of such fluctuations we
calculate the Fourier transform of the detection rate, at first for
the run with the Lattimer and Swesty EoS. To simplify this process
we have resampled the data in exact 1~ms intervals over a range of
400~ms, discarding a few ms of data at the beginning and end of the
original sequence that covered the interval 10.5~ms to 417~ms post
bounce. We take the Fourier transform on the full 400~ms interval,
however applying a Hann window function to reduce edge effects
(Appendix~\ref{sec:shotnoise}). From Fig.~\ref{fig:5wedge} it is
apparent that fast time variations commence in earnest at about
150~ms postbounce and we could have left out the initial phase of
the signal to reduce the number of background events within the
signal region. On the other hand, the window function anyway
suppresses the signal portion at the edges and so we have kept the
full 400~ms range.

The power spectra are in absolute units, not relative to the average
signal. In our normalization the Fourier amplitude at $f=0$ is
$N_{\rm events}/N_{\rm bins}$, i.e.\ the average rate per bin
(Appendix~\ref{sec:shotnoise}). In this way the average power does
not change if we consider sub--samples of the data for a shorter
duration, keeping the individual bin width fixed. The power at $f=0$
is therefore $(N_{\rm events}/N_{\rm bins})^2$. Since a typical
event rate is 1000~ms$^{-1}$ and our bin width is 1~ms, the power
spectrum at $f=0$ is around $10^6$ and much smaller at other
frequencies. Using a 400~ms time interval implies that the natural
frequency spacing is 2.5~Hz, and based on our 1~ms binning the
largest frequency that can be resolved is 500~Hz (Nyquist
frequency).

Based on the background rate of Eq.~(\ref{eq:backgroundrate}), a
signal duration of $\tau=400$~ms and a Hann window function, the
shot noise power of the detector dark current is given in
Eq.~(\ref{eq:shotnoise}) and is found to be 10.08 in the described
units where the zero-frequency power is $(N_{\rm events}/N_{\rm
bins})^2$. In the subsequent plots this power level is shown as a
horizontal grey line.

The modification of our power spectra plots as a function of SN
distance $D$ is not entirely trivial. If the SN is further away than
10~kpc, we lose event rate quadratically with distance. Since we are
showing the power spectrum, another power of 2 arises. Therefore,
relative to the fixed IceCube dark current (the horizontal line),
the power spectra are lowered by a factor $(10~{\rm kpc}/D)^4$. The
distance distribution of galactic SNe is very broad, but declines
quickly beyond about 20~kpc \cite{Mirizzi:2006xx}. At this
pessimistic distance, the signal power spectrum would be lowered by
a factor of 16, whereas the shot noise level would remain fixed.

On the other hand, if the SN is closer than 10~kpc, the IceCube dark
current quickly becomes irrelevant. The shot noise is determined
by the number of detected SN events which increases with decreasing
distance as $(10~{\rm kpc}/D)^2$ and therefore the shot-noise level
increases with this factor. The signal power spectrum increases with
the fourth power $(10~{\rm kpc}/D)^4$ as before, so relative to the
shot-noise level the signal power increases quadratically with
decreasing distance as $(10~{\rm kpc}/D)^2$.

\begin{figure}[t!]
\includegraphics[width=\columnwidth]{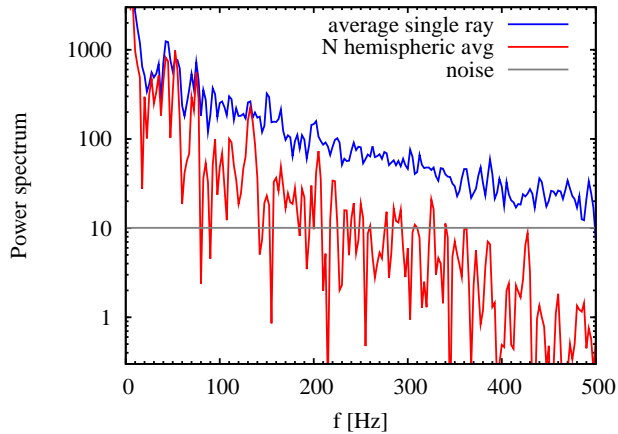}
\caption{\label{fig:suppression1} Power spectrum of the IceCube event
  rate for our model (EoS of Lattimer and Swesty)
  in 2.5~Hz increments.  Blue line (upper curve):
  Average single-ray power spectrum. Red line (lower
  curve): northern hemispheric average. Horizontal line: Shot noise
  from IceCube dark current.}
\end{figure}

\subsection{Hemispheric averaging}

As a first example we show in Fig.~\ref{fig:suppression1} as a blue
line (upper curve) the power spectrum of the event rate based on a
single ray. To this end we have taken an ensemble average of the
power spectra of all 96 angular rays with equal weights, treating
each one as if it were responsible for the full 4$\pi$ neutrino
emission and thus for the full detector signal. Taking a true single
ray instead of an ensemble average shows the same trend with much
greater noise and of course with directional differences. (Such a
single-ray treatment would correspond to the assumption that all
neutrinos are emitted strictly in the radial direction and the
observer receives neutrinos just from one spot on the stellar
surface.) Up to the Nyquist frequency of 500 Hz the single-ray power
stays far above the background and thus would be clearly detectable.

\begin{figure}[t!]
\includegraphics[width=\columnwidth]{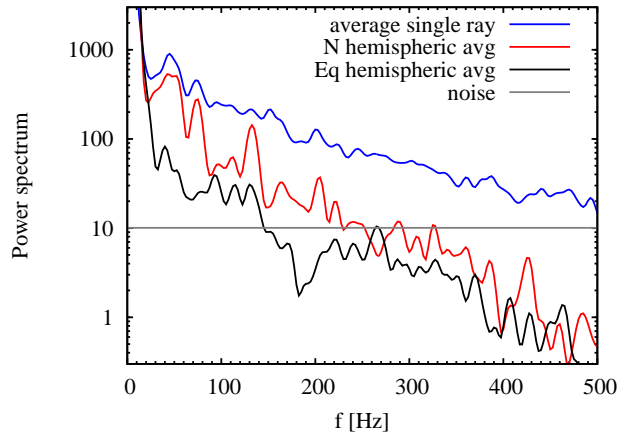}
\caption{\label{fig:suppression2}
Same as Fig.~\ref{fig:suppression1} with smoothing by a moving average
with a Gaussian window function of width $\sigma=4$~Hz.
Average power of all 96 single rays (blue),
northern hemispheric average (red), and equatorial hemispheric
average (black), from top to bottom.}
\end{figure}

The true signal is caused by the integrated emission over the
hemisphere facing the detector. In principle, it could be
reconstructed from the detailed angular information of the neutrino
emission as provided by our numerical solution of the neutrino
transport. However, in view of the approximative nature of 2D models
and of the ``ray-by-ray plus'' transport treatment (see Sec.~III.A)
we preferred to avoid processing the huge amount of corresponding
data. As a simplification we assumed that each surface element
radiates isotropically with an energy spectrum according to the
local conditions. In other words, we added the computed local
fluxes, weighted by the projected area of the surface elements as
seen by the observer. Taking the northern hemispheric average in
this sense leads to the red (lower) curve in
Fig.~\ref{fig:suppression1}.

The spectral power declines much faster with increasing frequency
than in the single-ray case. This is expected because high-frequency
luminosity variations are due to small spatial scales of the
``boiling'' medium whereas the lowest frequencies are due to the
largest-scale convective overturns and SASI activity. The variations
on small spatial scales are not strongly correlated and therefore
reduced when taking an average over the entire hemisphere whereas
the large-scale motions are correlated and not averaged away.

The hemispheric power spectrum is quite noisy and the overall trend
is better seen in a smoothed version shown in
Fig.~\ref{fig:suppression2}. Here we show the same information based
on a moving average with a Gaussian window function with
$\sigma=4$~Hz. In addition we show the equatorial average as a black
curve. It is the lowest curve and has significantly less power than
the northern case (which is similar to the southern one). In this
simulation the large-scale motion is essentially along the symmetry
axis of the simulation, explaining much larger luminosity variations
in the polar directions than the equatorial one. Indeed, the
luminosity and temperature variations in the north and south are
anti-correlated because of the dipole nature of the SASI
oscillation, so in the equatorial view the variations largely
cancel. In a realistic 3-D situation, the dipole direction is not
necessarily fixed in space, so over several hundred ms the average
view from different directions probably would not differ as
dramatically as in this axisymmetric simulation.

\subsection{Comparison with spherically symmetric case}

\begin{figure}
\includegraphics[width=\columnwidth]{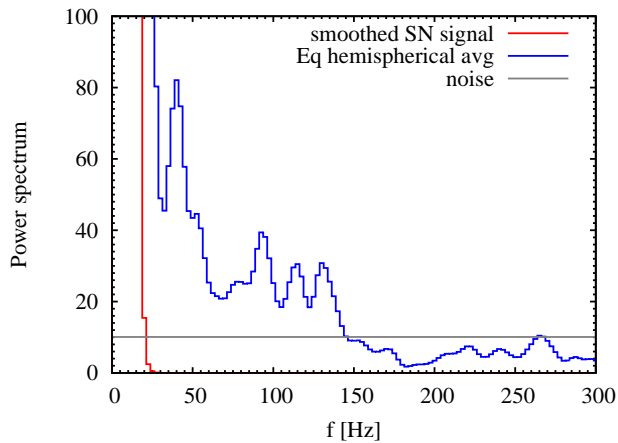}
\caption{\label{fig:sasiVSnosasi} Comparison between fluctuating and
  smooth signal for the model with Lattimer and Swesty EoS. 
  Blue: Power spectrum of equatorial hemispheric
  average signal. Red: Same for smoothed SN signal.}
\end{figure}

An important issue is how well one can distinguish the signal from a
spherically symmetric model from a convecting one. To this end we
have produced an equivalent spherically symmetric model by smoothing
the output from our model by a moving average. An example for the
corresponding smooth luminosity is shown in
Fig.~\ref{fig:firstexample}. We compare the signal power spectrum
for the equatorial hemispheric average in
Fig.~\ref{fig:sasiVSnosasi} with that from a smoothed version of
this average. We see that the smooth signal plummets below the
IceCube background noise level at around 20~Hz. In other words, for
the case studied here the power spectrum roughly above 20~Hz is a
clear indication for fast time fluctuations of the neutrino~source.

\subsection{Directional and flavor dependence}

\begin{figure}
\begin{tabular}{c}
\resizebox{80mm}{!}{\includegraphics{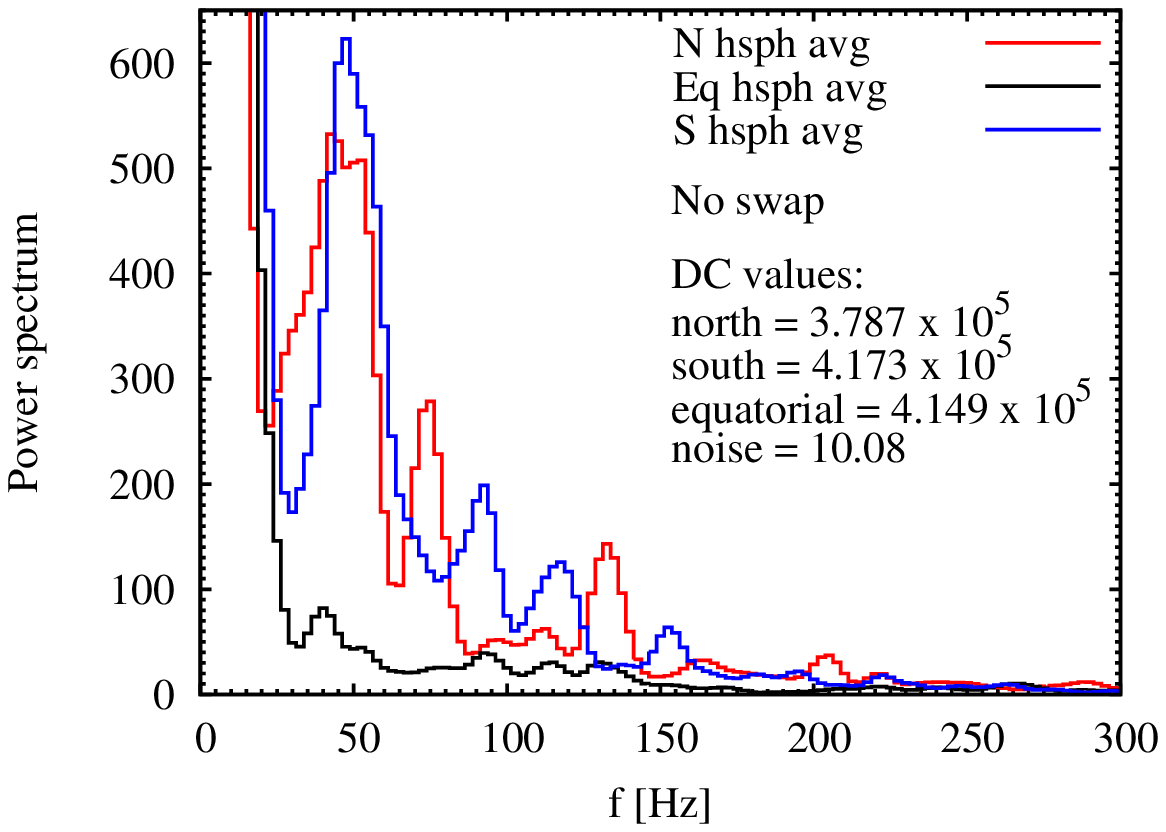}}\\
\resizebox{80mm}{!}{\includegraphics{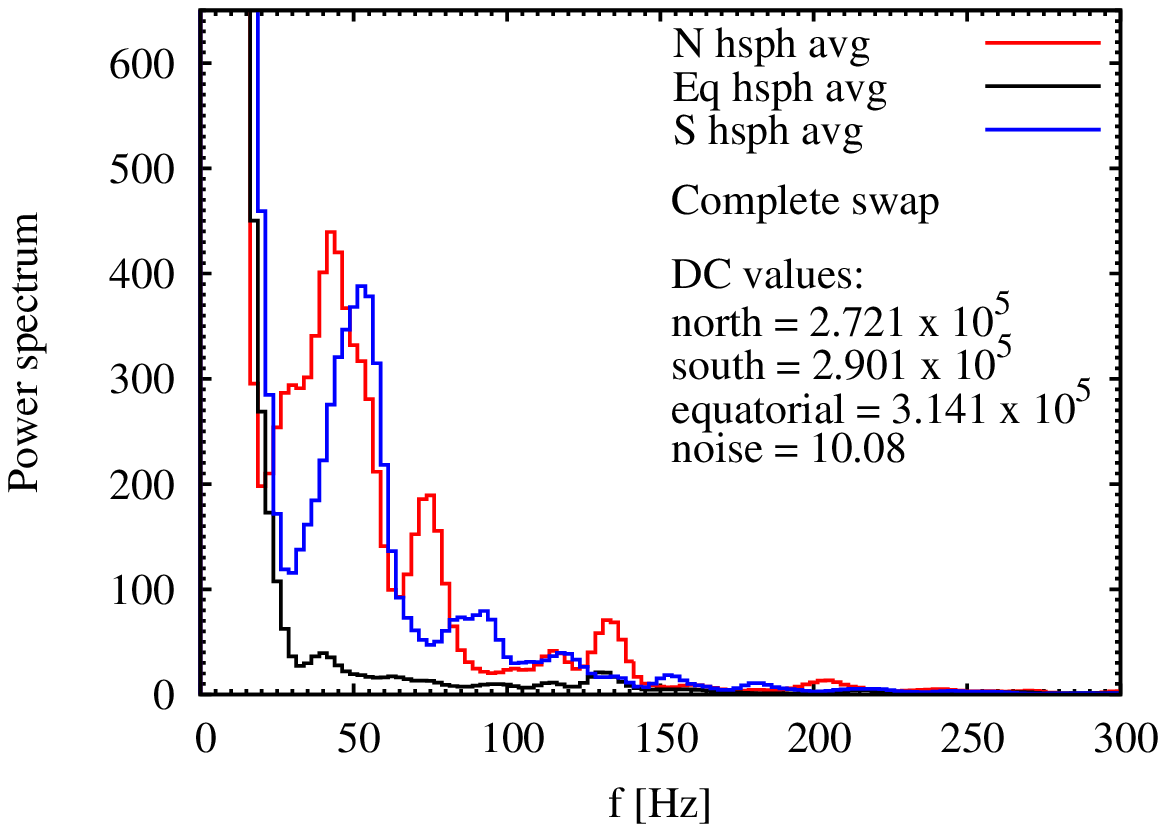}}
\end{tabular}
\caption{\label{fig:pwl_anti}Smoothed power spectrum of the event rate
  for the northern (red), equatorial (black) and southern (blue)
  hemispheric averages (EoS of Lattimer and Swesty).
  The shot noise level of the background is at 10.08.
  {\it Top:} No flavor oscillations. {\it Bottom:} Complete swap
  $\bar\nu_e\leftrightarrow\bar\nu_x$.
  In the panels we also give the DC values, i.e.\ the power
  at zero frequency, corresponding to
  $(N_{\rm events}/N_{\rm bins})^2$.}
\end{figure}

For a more detailed appreciation we next show in
Fig.~\ref{fig:pwl_anti}, upper panel, the smoothed power spectrum
for the northern, southern and equatorial average signals. We first
observe that from $\sim$20~Hz to $\sim$175~Hz for $\bar{\nu}_e$ the
power spectrum for all three hemispheric averages are comfortably
above the noise level in IceCube.  We furthermore see that the first
peak for both polar directions are roughly coinciding, and although
the specific pattern differs at larger frequencies the levels are
comparable. The pronounced peak at 50~Hz corresponds to variations
with a 20~ms period. This period is easily seen in Fig.~5 of
Ref.~\cite{Marek:2008qi} where the dipole motion of the shock-wave
surface is plotted.

In the lower panel we show the same signal under the assumption that
complete flavor transformations have taken place and what reaches
IceCube are $\bar\nu_e$ that at the SN were born as $\bar\nu_x$. The
qualitative features are similar as before. In other words, full or
partial flavor transformations would not change the picture
substantially. The overall power spectrum is now slightly lower.
However, this effect is due to the reduced $\bar\nu_x$ luminosity
during the accretion phase relative to the $\bar\nu_e$ luminosity.
In the figure panels we give the DC (``direct current'')
values of the power spectrum,
i.e.\ the power at zero frequency which is significantly larger in
the $\bar\nu_e$ case (upper panel). In other words, the {\em
relative\/} fluctuation amplitude is similar for both species.
\begin{figure}
\resizebox{80mm}{!}{\includegraphics{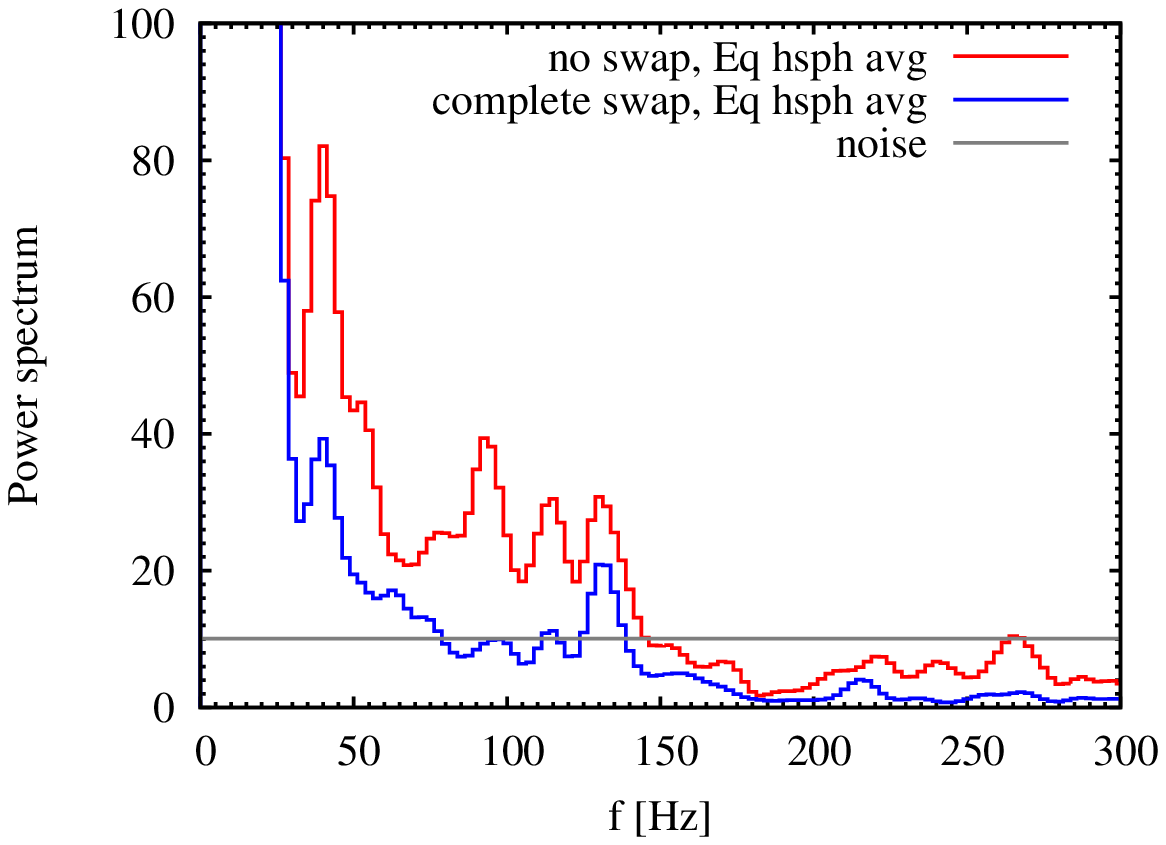}}\\
\resizebox{80mm}{!}{\includegraphics{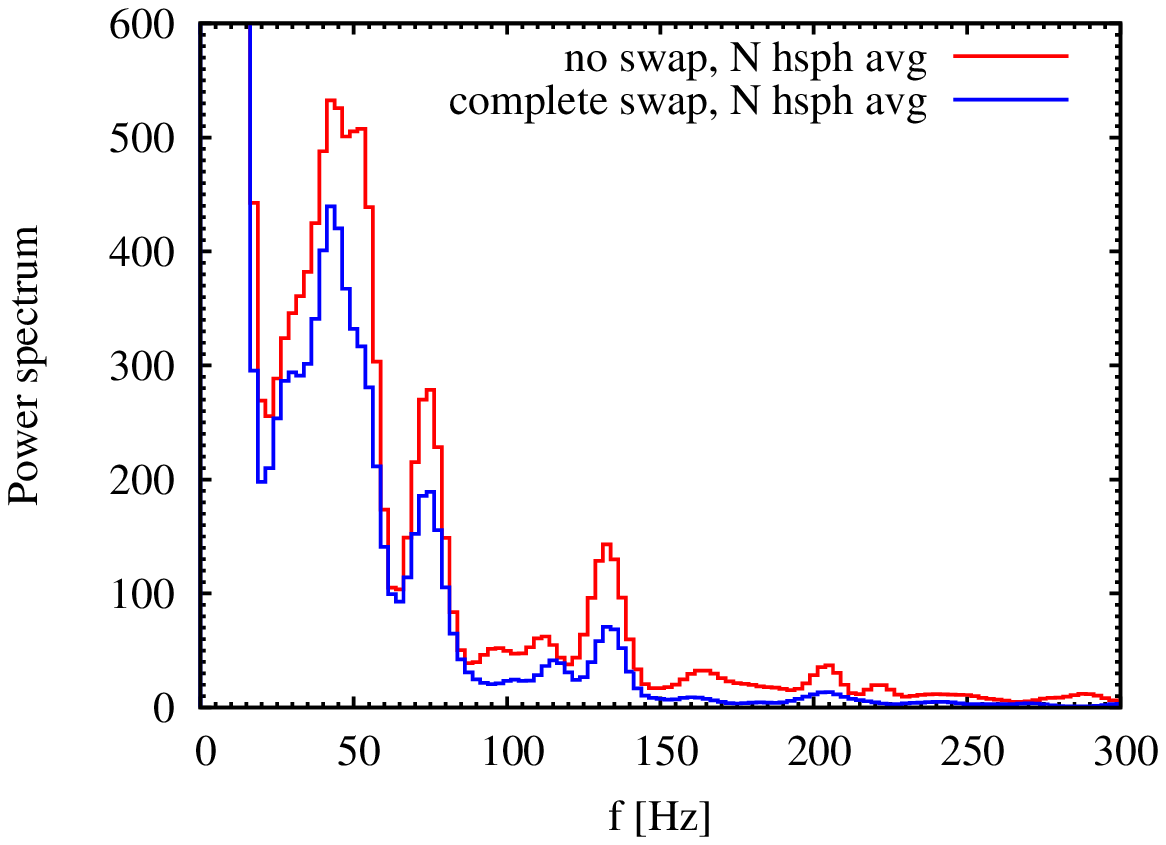}}
\caption{\label{fig:flavorswap}Comparing no oscillations (red) with
  complete flavor swap (blue) for the model with Lattimer and Swesty
  EoS.
  {\it Top:} equatorial. {\it Bottom:} northern.\label{fig:flavors}}
\end{figure}

The same information is given in Fig.~\ref{fig:flavors} where we
show the equatorial average (top panel) and northern average
(bottom) and each time compare the $\bar\nu_e$ signal with the case
of complete flavor swap. Especially the northern case (very similar
to the southern one) shows that flavor transformations have little
impact on the interpretation of fast signal variations.

\section{Luminosity vs.\ energy fluctuations}
\label{sec:correlations}

Thus far we have focussed on the counting-rate fluctuations in
IceCube because among existing detectors it provides by far the
largest event rate. With a future megaton-class water Cherenkov
detector the picture would change because the event rate would be
comparable to IceCube and in addition one would obtain
event-by-event energy information. In this case spectral
fluctuations would become important as well.

\begin{figure}[b]
\resizebox{80mm}{!}{\includegraphics{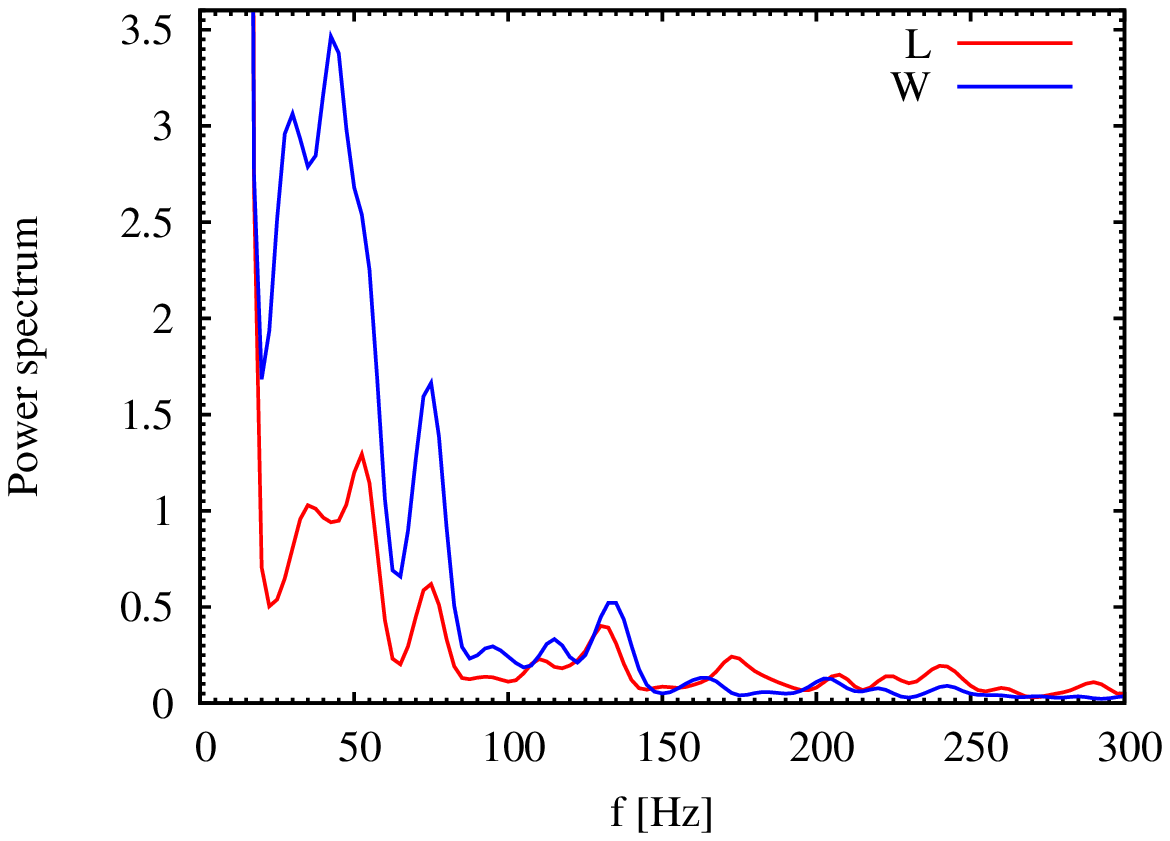}}
\resizebox{80mm}{!}{\includegraphics{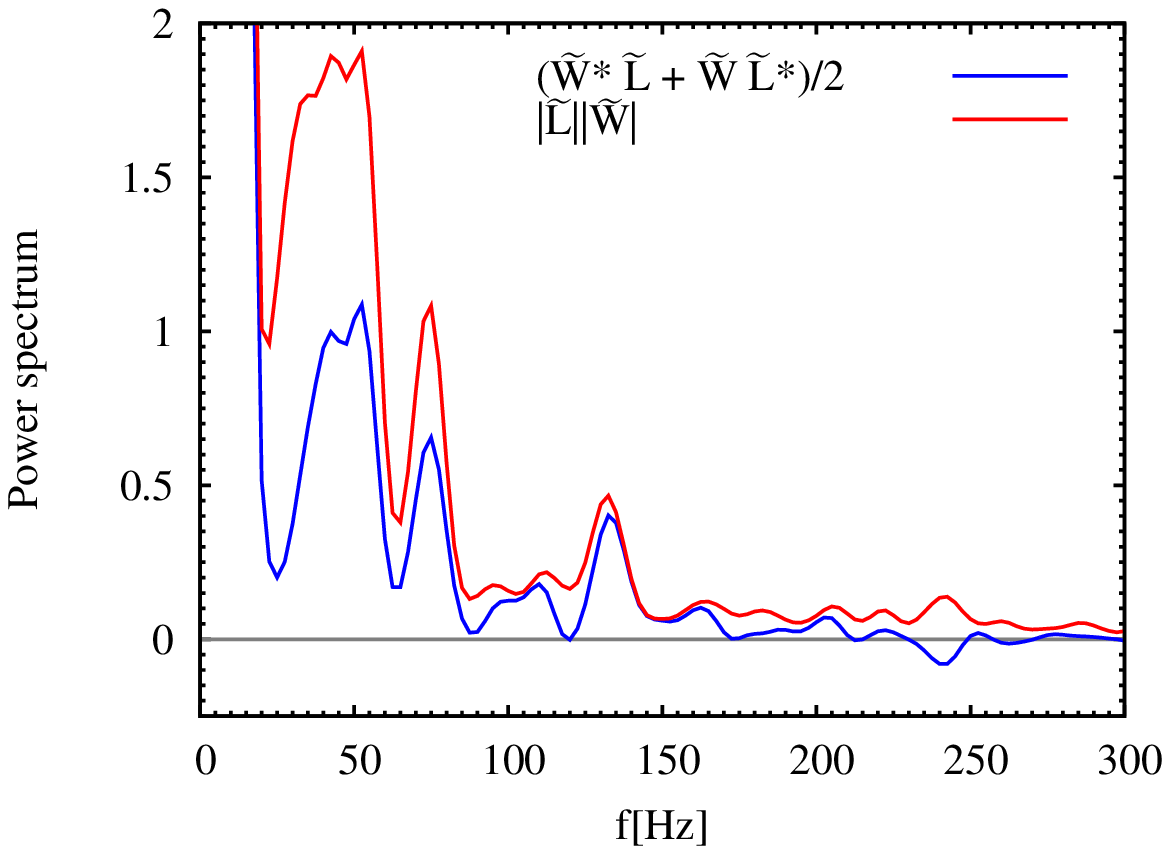}}
\caption{\label{fig:correl}Smoothed power spectrum of luminosity ($L$) 
and spectral energy ($W=E^2_{\rm rms}$) fluctuations for the
model with Lattimer and Swesty EoS. $L$ and $W$ were
normalized to their respective average values over the 274--400~ms
interval. {\it Top:} Power spectra of $L$ (red, lower curve) and
$W$ (blue, higher curve). {\it Bottom:} $|\tilde L|\,|\tilde W|$ as a
red line and correlation function $\frac{1}{2}(\tilde W^* \tilde L+
\tilde W \tilde L^*)$ as a blue line.}
\end{figure}

In Fig.~\ref{fig:firstexample} it is apparent that the IceCube
signal variations are much larger than the luminosity variations.
The fluctuations of $E_{\rm rms}$ must be responsible for the
difference. Moreover, one expects that the spectral fluctuations are
correlated with the luminosity fluctuations so that both effects
interfere constructively.

To quantify these arguments we consider as a specific example the
northern hemispheric average of the luminosity $L(t)$ and the rms
energy $E_{\rm rms}(t)$ and their correlation. Since we are here
concerned with the IceCube signal, that is proportional to $E_{\rm
  rms}^2$, we consider the two functions $L(t)$ and $W(t)=E_{\rm
  rms}^2(t)$. For convenience we normalize them somewhat arbitrarily
to their average values over the 274--400~ms interval since now we are
primarily interested in relative fluctuations. (Our conclusions do not
change much if we normalize to the average values over the entire
400~ms interval.) We next calculate the Fourier transforms $\tilde L$
and $\tilde W$ of these dimensionless functions. In the upper panel of
Fig.~\ref{fig:correl} we show the power spectra. The spectral power of
$W$ is significantly larger than that of $L$, i.e.\ the IceCube signal
variations are dominated by $E_{\rm rms}^2$ variations.

To quantify correlations between spectral and luminosity variations
we show in the lower panel of Fig.~\ref{fig:correl} the quantities
$|\tilde W|\,|\tilde L|$ and $\frac{1}{2}(\tilde W^* \tilde L+
\tilde W \tilde L^*)$. The two quantities are similar and the
correlation function is positive almost everywhere, so indeed $W(t)$
and $L(t)$ are strongly correlated.

These results suggest that energy- and event-rate fluctuations and
their correlation, that could be measured in a megaton water
Cherenkov detector, would provide additional signatures for
convection and SASI activity.

\section{Stiff Equation of State}                     \label{sec:HWeos}

Finally we briefly address the dependence of our results on the EoS
used in the SN simulation. To this end we consider the run of
Ref.~\cite{Marek:2008qi} with the EoS of Hillebrandt and
Wolff~\cite{Hillebrandt:1985}. The $\bar\nu_e$ luminosity, rms
energy and IceCube rate corresponding to our five angular wedges was
shown in Fig.~\ref{fig:5wedgeHW} in juxtaposition to the run
with the Lattimer and Swesty EoS. In Fig.~\ref{fig:firstexampleX} we
showed the $\bar\nu_e$ luminosity, rms energy and IceCube signal rate for the
northern hemispheric average with the Hillebrandt and Wolff EoS in analogy to
Fig.~\ref{fig:firstexample}.

\begin{figure}
\includegraphics[width=\columnwidth]{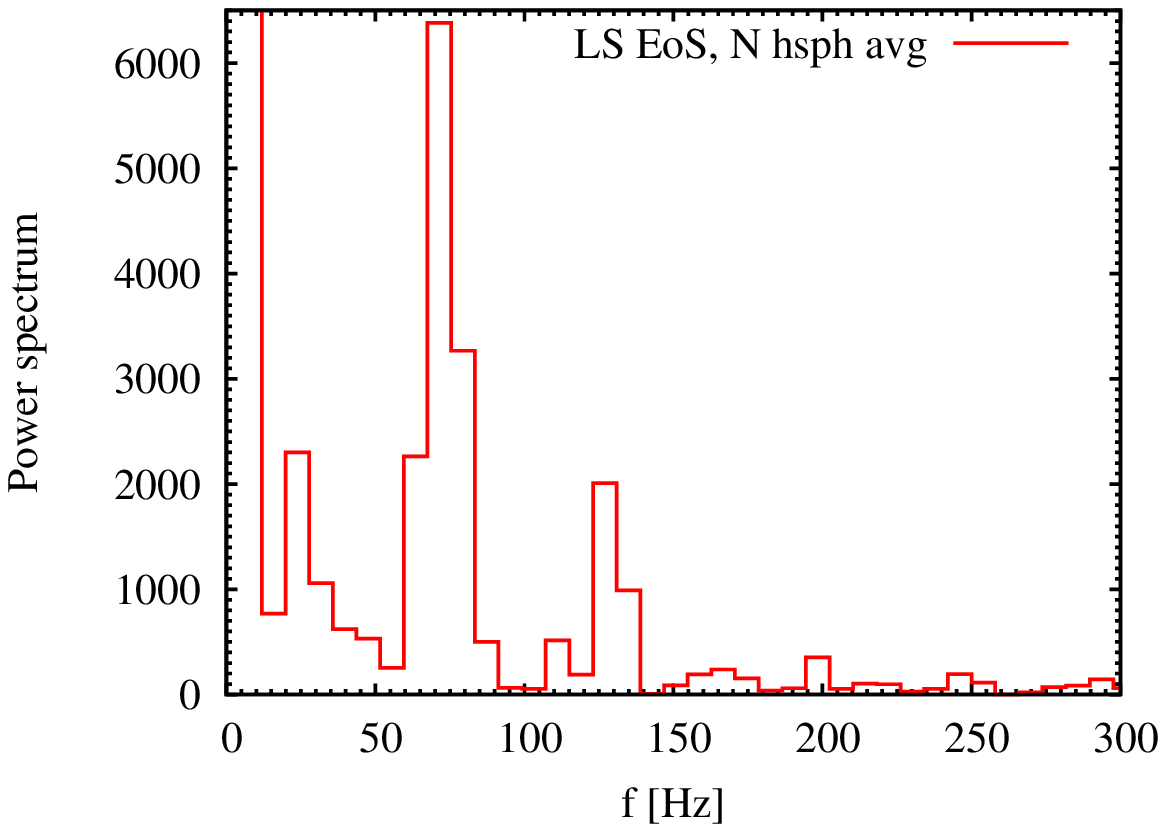}
\includegraphics[width=\columnwidth]{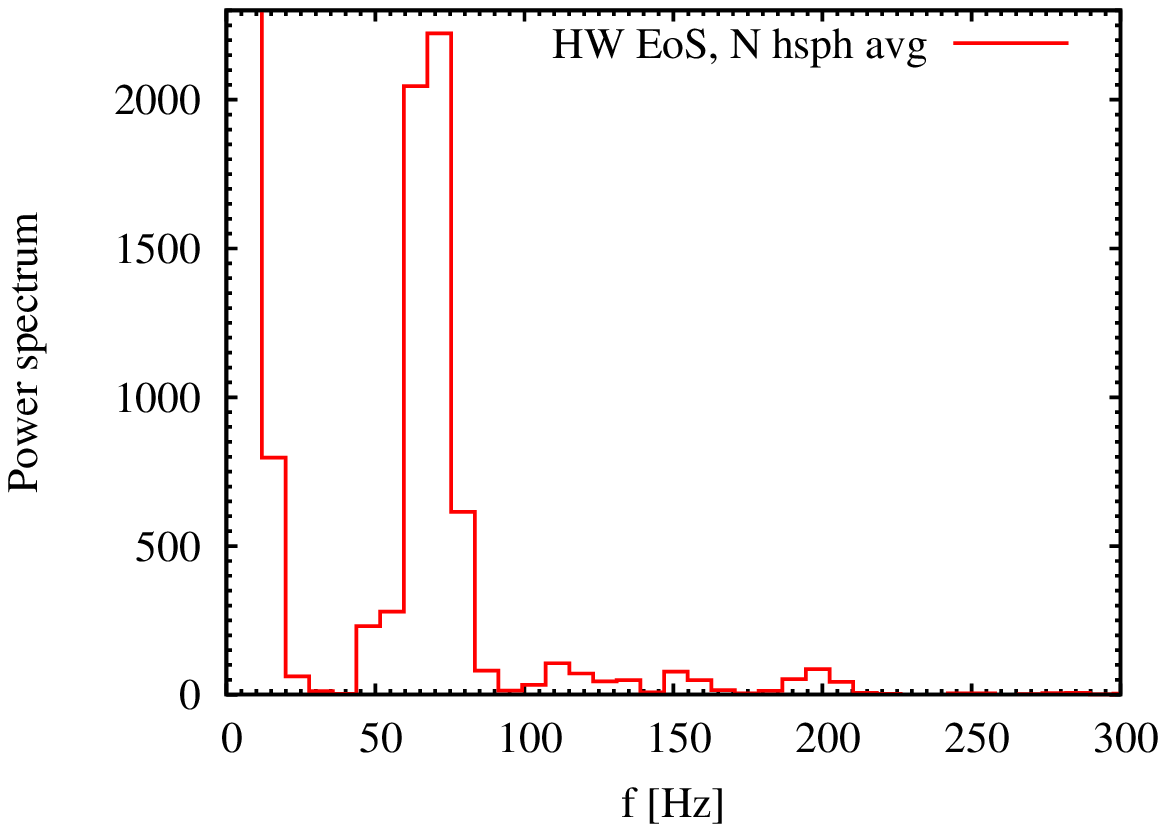}
\caption{\label{fig:compareEoS}Power spectra of the IceCube rate without
flavor oscillations for the northern
hemispheric averages, taken for the 274--400~ms post bounce interval.
Because of the reduced integration time,
the shot noise level of the IceCube dark current is now at 30.26.
{\it Top:} Lattimer and Swesty EoS in analogy to Fig.~\ref{fig:pwl_anti}.
{\it Bottom:} Hillebrandt and Wolff EoS.}
\end{figure}

The figures reveal that strong fluctuations begin in earnest only at
around 300~ms postbounce as already noted in
Ref.~\cite{Marek:2008qi}. Therefore, it makes little sense to
compare the power spectrum over the first 400~ms of this run with
the Lattimer and Swesty case. Instead we compare in
Fig.~\ref{fig:compareEoS} the power spectra of the IceCube rate for
the northern hemispheric averages, taken over the 274--400~ms
interval post bounce for the Lattimer and Swesty case (top panel)
and the Hillebrandt and Wolff case (bottom). In other words, our
signal duration is now $\tau=126$~ms, implying a frequency spacing
of 8~Hz and an increase of the shot-noise level by a factor
$400/126=3.2$ (Appendix~\ref{sec:shotnoise}).

Both cases show strong power with a frequency of around 70~Hz which
is also visible to the naked eye in the time evolution (lower panel
of Fig.~\ref{fig:firstexampleX}), although the power is larger in
the Lattimer and Swesty case. The signal time variations for the
stiffer EoS would be plainly visible with similar significance as for
the softer one.

We also note that the 70~Hz peak in the upper panel of
Fig.~\ref{fig:compareEoS} is much larger than in
Fig.~\ref{fig:pwl_anti} (upper panel, red line). The difference
between the two curves is only the analyzed signal interval. Here it
is the final 126~ms of the run, in Fig.~\ref{fig:pwl_anti} the full
400~ms. If the signal was roughly stationary, the power of the peak
would have to be the same. However, here we have a strong peak at
70~Hz with a much larger signal-to-noise than in
Fig.~\ref{fig:pwl_anti}, where in turn we have a much stronger peak
at 50~Hz. In other words, the Fourier spectrum varies significantly
as a function of time. The analysis of a realistic signal would
involve studying subsets of the full-length signal where Fourier
components can show up with much larger significance in spite of the
increased shot noise relevant for a shorter integration time.

\section{Implications of detection}           \label{sec:implications}

Hydrodynamical instabilities, in particular convective and SASI
activity, and multi-dimensional processes are thought to be crucial
ingredients of the mechanism that causes the explosion of
core-collapse SNe \cite{Janka:2007, Burrows:2007a}. The nonradial
asymmetries during the very early stages of the explosion can
manifest themselves in large-scale asphericity and mixing of the SN
blast (for recent 3-D models see Ref.~\cite{Hammer:2009cn}). A
measurement of neutrino signal modulations and gravitational
waves would provide direct evidence for these theoretical ideas and
could yield much deeper insight into the strength, evolution, and
role of nonradial hydrodynamic flows on the path to successful
explosions. While the main focus of this paper is the experimental
detectability of SASI and convective variations of the neutrino
emission---at least as predicted by 2-D simulations---we briefly
address possible interpretations of such a measurement.

A thorough discussion is hampered by the small number of SN models
that are available for an analysis of the signal characteristics and
dependence on the progenitor and core microphysics. Moreover, a
conclusive theoretical assessment will require 3-D models. The
present 2-D calculations can serve only for preliminary indications
of what might be expected, provided 3-D models roughly confirm the
2-D results. The prominence of strong dipolar asymmetries as
obtained by the lowest ($\ell = 1$ in terms of an expansion in
spherical harmonics) SASI modes in 2-D has indeed been questioned on
the basis of recent 3-D simulations \cite{Nordhaus:2010}. However,
these models do not include neutrino transport in a self-consistent
way and only a small set of non-rotating models for special
conditions was computed. Therefore, it is premature to judge the
role of low-mode asymmetries in 3-D.

The peaks of the power spectrum of the neutrino signal reflect the
mode pattern of the SN core activity. They are connected with
temporal changes of the mass accretion rate onto the neutron star,
which are either caused by SASI or by convective modulations of the
infalling mass flow between shock and neutron star
surface~\cite{Marek:2007gr, Marek:2008qi}. Since the accretion
downflows can be channeled more strongly to the northern or southern
hemisphere, the hemispheric symmetry can easily be broken, and even
long-lasting hemispheric asymmetries of the accretion structures and
shock-expansion strength can emerge. Therefore there is no reason to
expect exactly equal power spectra for both hemispheres. Even the
dipolar SASI sloshing mode can develop different strengths in both
hemispheres.

A causal connection of the peaks in the neutrino power spectra
with global deformation modes of the shock surface and thus of the 
accretion flow between shock and proto-neutron star is supported by
results presented in Ref.~\cite{Marek:2008qi}. The Fourier spectra
of the time-variable spherical harmonics dipole and quadrupole
amplitudes of the shock 
position in that paper (Fig.~5, right column) exhibit maxima whose
positions agree well with the lowest characteristic frequencies of 
the neutrino power spectra. 

The corresponding 50$\,$Hz peak of our model with Lattimer and
Swesty EoS is actually a broad feature with
high power (half-width) roughly between 30$\,$Hz and 60$\,$Hz. This
feature is present in the northern, southern, and equatorial
signals. In the equatorial case we identify it with the first
two-hump maximum visible to the right of the low-frequency spike
(Fig.~\ref{fig:pwl_anti}, upper panel). This peak around 50$\,$Hz is
caused by the $\ell = 1$ SASI sloshing mode, which leads to
quasi-periodic modulations of the mass accretion rate and associated
neutrino emission in both hemispheres (see Ref.~\cite{Marek:2008qi},
page 485). The peak width is explained by the time-variations of the
SASI frequency, its north-south differences by the lack of perfect
hemispheric symmetry (cf.\ Fig.~3 in Ref.~\cite{Marek:2008qi}).

The frequency of the $\ell = 1$ SASI mode depends on the
sound-travel and mass-inflow times between neutron star surface and
shock (see Eq.~(32) in~\cite{Ohnishi:2005cv} and Eq.~(18)
in~\cite{Scheck:2008}) and thus mainly on the time-variable shock
radius and to a lesser extent also on the continuously contracting
NS radius. Roughly, when the average shock radius is large, the SASI
frequency is lower, when the shock radius shrinks, the SASI
frequency tends to be higher. Since the shock radius shows sizable
time evolution (Fig.~4, left panel, in \cite{Marek:2008qi}), it is
natural that the SASI peak of the time-integrated neutrino signal
becomes fairly broad.

The peak with the next higher frequency of around 70$\,$Hz is most
easily explained by $\ell=2$ (quadrupolar) SASI activity, although a
strict discrimination of global shock oscillation modes from
convective mass motions in the postshock layer 
is very difficult in the nonlinear phases of the two hydrodynamic
instabilities. Both instabilities can trigger each other and
therefore occur mostly simultaneously \cite{Scheck:2008}. A possible
connection of the $\sim$70$\,$Hz power maximum with the quadrupolar
SASI mode is suggested by three facts: (1) Analytic analysis and
numerical experiments in the linear regime show that the frequency
of the $\ell=2$ mode is slightly higher than that of the $\ell=1$
mode (cf.\ Fig.~5 in \cite{Ohnishi:2005cv}, Fig.~13 in
\cite{Scheck:2008}), but the exact frequency ratio depends strongly
on the size of the SASI region (see Fig.~4 in \cite{Blondin:2006}).
(2) The appearance of the $\sim$70$\,$Hz peak in the upper panel of
Fig.~\ref{fig:compareEoS} and relative weakness of power at lower
frequencies at late times (274--400$\,$ms) in our model with the
Lattimer and Swesty EoS can be explained by the increasing power of
the $\ell=2$ shock oscillation mode, whose amplitude at these times
becomes larger than that of the $\ell=1$ mode (see the left panels
of Fig.~5 in \cite{Marek:2008qi}). (3) The presence of the peak in
the northern power spectrum but relative weakness or absence in the
southern hemisphere may correspond to the hemispheric asymmetry of
the quadrupolar shock deformation as visible in the panels of Fig.~4
in Ref.~\cite{Marek:2008qi}.

Local convective overturn motions in the accretion flow on smaller 
angular scales (corresponding to higher spherical harmonics modes)
take place on shorter timescales than the global dipolar and quadrupolar
mass shifts. Therefore they are the most probable explanation for
the power peaks at frequences above 90--100$\,$Hz. In particular
such short-wavelength structures may exhibit considerable differences
in the two hemispheres and also in the equatorial region, where
long-lasting, non-stationary downdrafts develop at later post-bounce 
times. Differences between these directions in the neutrino power 
spectra at high frequencies are therefore not astonishing.

Moreover, the layer between shock and neutron star surface is
generally more compact for the SN model with the Hillebrandt and
Wolff EoS (cf.\ Fig.~4, left panel, of \cite{Marek:2008qi}).
Therefore, the global shock-motion and accretion modes of this model
have a lower amplitude and higher frequency than in the simulation
with the Lattimer and Swesty EoS. It is possible that the very broad
peak roughly between 60$\,$Hz and 85$\,$Hz in the former case (lower
panel of Fig.~\ref{fig:compareEoS}) and simultaneous lack of a
second strong peak at somewhat higher frequency is the result of a
superposition of $\ell=1$ and $\ell=2$ activity. This interpretation
is suggested by the presence of both modes with comparable
amplitudes (Fig.~5 in \cite{Marek:2008qi}) and by the fact that for
a compact postshock layer the characteristic frequencies of both
modes become very similar as shown in Fig.~4 of
Ref.~\cite{Blondin:2006}.

SASI activity can therefore be reflected by different peaks in the
neutrino power spectra, depending on the presence of different
modes, the fastest growing ones and typically strongest in 2-D being
those of $\ell = 1$ and \hbox{$\ell=2$}. The exact frequencies of
the peaks depend on the time-evolving structure of the postshock
layer. The strongest SASI activity and thus most easily measurable
signal features must be expected to occur in a time window of some
100$\,$ms just before the explosion sets in. The peaks in the
power spectra of the integrated signal over this shorter period are
significantly enhanced relative to the shot-noise level (compare
upper panels of Figs.~\ref{fig:pwl_anti} and \ref{fig:compareEoS}),
favoring easier detection of these features.

The measurement of the neutrino luminosity modulations could thus
confirm the existence of large non-radial hydrodynamical
instabilities (SASI and convective overturn) around the beginning of
the explosion, which would have to be strong enough to affect the
neutrino emission from the accreting neutron star. Such a
measurement would reveal an important component of SN physics, whose
potential relevance is presently suggested only by numerical models,
theoretical analysis, and indirect arguments based on the presence
of ejecta asymmetries at much later stages of the evolution.

A detection of the SASI would definitely exclude the prompt
explosion mechanism (which already seems to be ruled out by
simulations) as well as all other explosion mechanisms that work
faster than the SASI can develop in the SN core. The growth of the
SASI activity to the nonlinear regime takes typically 100--200 ms
after bounce \cite{Ohnishi:2005cv, Blondin:2006, Scheck:2008}, so
SASI signatures would require a significantly delayed explosion as
expected for the neutrino-driven mechanism. Most probably, their
measurement would also exclude the magnetohydrodynamic mechanism,
which could take place in rapidly rotating stellar cores and
according to 2-D simulations could lead to relatively rapid
explosions \cite{Burrows:2007c}. Conversely, a non-detection of
strong SASI features in the neutrino signal is likely to disfavor
the acoustic explosion mechanism \cite{Burrows:2006, Burrows:2007b},
which might initiate the blast wave as late as one second or more
after core bounce. In this case the onset of the explosion would be
preceded by at least a transient phase of strong SASI and convective
overturn activity around the new-born neutron star. Such a phase is
probably important to excite the compact remnant to the required
powerful, large-amplitude dipole oscillations that yield the
acoustic energy flux for launching the explosion.

Exclusion arguments of this kind can become even more powerful in
the combination with gravitational-wave measurements as recently
pointed out by Ott~\cite{Ott:2008wt}. Thus gravitational waves and
neutrino-emission variations could help to unravel the still heavily
disputed processes that cause the explosions of massive stars.

\section{Conclusions}                          \label{sec:conclusions}

We have studied the signature of fast SN neutrino time variations in
IceCube, a detector that would produce the largest event rate of any
existing experiment. We have used the output of axially symmetric SN
simulations recently produced by some of us. The SASI sloshing motion
with time- and mode-dependent frequencies of about 50--100~Hz as well
as smaller-scale, shorter-period convective overturns provide a strong
imprint in the neutrino signal. Typically it would be visible even to
the naked eye by simply inspecting the time sequence of registered
Cherenkov photons. A Fourier analysis of the signal reveals a large
signal-to-noise ratio that would be detectable for a SN throughout our
galaxy.

The spectral power of the time-varying SN signal decreases with
frequency and it depends on distance up to which frequency time
variations can be detected. For a fiducial distance of 10~kpc the
IceCube dark current is comparable to the SN signal.  Based on our
simulation, at this distance signal modulations typically could be
seen up to 100--200~Hz. This conclusion is barely affected by possible
flavor conversions.

The strongly dipolar nature of the SASI mode along the symmetry axis
of the simulations implies that the observable signal variations
strongly depend on the direction of viewing the SN. In particular, in
the equatorial direction the signal variations caused by neutrinos
emerging from the northern and southern hemispheres nearly cancel and
in our most pessimistic example would be visible only to a distance of
a few kpc. However, such directional cancelation effects likely would
be smaller in a realistic 3-D situation, although the overall SASI
signal might also be smaller.  Moreover, the signal-to-noise for a
given Fourier mode depends on the time window used for the analysis
because the power spectrum varies strongly with time. It is premature
to study these issues in too much detail because the available 2-D
simulations as well as the approximations used in the neutrino transport
provide only a first glance of what might be expected from a more
realistic treatment.

The event rate fluctuations in IceCube are caused by fluctuations of
the luminosity and of the neutrino energies, the latter being the
more important effect. Therefore, in a detector with spectral
information such as a water Cherenkov detector additional
information can be extracted. A future megaton-class detector will
have a neutrino event rate comparable to IceCube's and thus would
offer significant additional capabilities through its event-by-event
spectral sensitivity.

Our main message is that IceCube and future large-scale detectors can
measure intriguing time-dependent features of the neutrino signal of a
future galactic SN, allowing one to observe the SASI activity with
neutrinos in situ, if our 2-D model is roughly representative for a
more realistic treatment. Such signatures would provide a crucial test
of our theoretical understanding of the core collapse phenomenon. The
secular evolution of the signal as well as its fast variations may
hold information, for example on the growth time of large-scale
non-radial asymmetries in the SN core, the SN explosion mechanism, and
the contraction behavior of the nascent NS and thus on the nuclear
equation of state, but once more it is premature to forecast generic
signatures on the basis of our 2-D models. Moreover, the spectral
range that can be probed strongly depends on the SN distance---a
fiducial case at 10~kpc may not be representative if the SN is much
closer or much further away.

The excellent time resolution of IceCube can be used in other
ways. For example, the signal onset and therefore bounce time can be
pinned down very well, allowing for correlations with gravitational
wave detectors~\cite{Halzen:2009sm}. On the more exotic side, a
possible QCD phase transition can produce a short $\bar\nu_e$ burst
that could be detected with high significance~\cite{Dasgupta:2009yj}.

In summary, among existing SN detectors IceCube has unique
capabilities to measure fast signal variations. Identifying such
features with additional spectral information is a powerful
motivation to build a megaton water Cherenkov detector.

\section*{Acknowledgments}

We acknowledge partial support by the DFG (Germany) under grant
TR-27 ``Neutrinos and Beyond,'' the Cluster of Excellence ``Origin
and Structure of the Universe,'' and the NSF under Grant No.\
PHY-0854827. We acknowledge computer time grants at the John von
Neumann Institute for Computing (NIC) in J\"ulich, the
H\"ochst\-leistungs\-re\-chen\-zentrum of the Stuttgart University
(HLRS) under grant number SuperN/12758, the
Leib\-niz-Re\-chen\-zentrum M\"unchen, and the RZG in Garching.
T.L.\ thanks the MPI Physics for hospitality while this work was
begun. We thank the participants of the workshop JIGSAW 2010 (22--26
February 2010, Mumbai, India) for comments and discussions, in
particular Timo Griesel and Thomas Kowarik.

\appendix

\section{Shot noise in IceCube and Fourier Transform of binned data}
                          \label{sec:shotnoise}

To estimate the shot noise of the IceCube signal we consider a
signal consisting of a sequence $t_j$ of $N$ measured arrival times.
They sample the rate $R(t)$ over the signal duration $\tau$. The
Fourier transform is
\begin{equation}\label{eq:FourierdefinitionAppendix}
g(f)=\int_0^{\tau}\D t\,R(t)\,\E^{-\I 2\pi f t}
=\sum_{j=1}^{N}\E^{-\I 2\pi f t_j}
\end{equation}
with the spectral power $G(f)=|g(f)|^2$. The detection rate has
units of inverse time, so the Fourier components are dimensionless
and $g(0)=N$ and $G(0)=N^2$. The finite-time Fourier transform is
limited to the discrete frequencies $f_k=k \Delta f=k/\tau$.  It is
understood that a frequency $f$ stands for a member of this discrete
set.

If the sequence of events samples a uniform distribution on the
interval $0\leq t_j\leq \tau$, the sum in
Eq.~(\ref{eq:FourierdefinitionAppendix}) represents a random walk in the
complex plane with unit step size. One concludes that an ensemble
average for $G(f)$ is independent of frequency for $f\not=0$ and
follows the normalized distribution $p(G)=N^{-1}\,\E^{-G/N}$
\cite{Dighe:2003jg}. The average is $\langle G\rangle=N$ so that
\begin{equation}\label{eq:1overNAppendix}
\frac{\langle G_{f\not=0}\rangle}{G_{f=0}}=\frac{1}{N}\,.
\end{equation}

Usually we will include a window function $w(t)$ on the interval
$0\leq t\leq\tau$ to suppress edge effects on the Fourier transform.
Therefore, we actually use
\begin{equation}
g(f)=\int_0^{\tau}\D t\,w(t)\,R(t)\,\E^{-\I 2\pi f t}\,.
\end{equation}
The average weight must be unity, implying
\begin{equation}
\int_0^{\tau}\frac{\D t}{\tau}\,w(t)=1\,.
\end{equation}
Fourier components that vary fast on the scale $\tau$ are returned
with their original amplitude.

We determine the impact of a window function on the shot noise by
extending the picture of a random walk in the complex plane to a
variable step size. The different random walks must be combined in
quadrature and the expectation value is modified as
\begin{equation}\label{eq:w2overN}
\langle G\rangle = \langle w^2\rangle\,N
\end{equation}
where
\begin{equation}
\langle w^2\rangle = \int_0^\tau\frac{\D t}{\tau}\, w^2(t)\,.
\end{equation}
We will specifically use the Hann window
\begin{equation}
w(t)=1-\cos(2\pi\,t/\tau)\,,
\end{equation}
implying that the shot-noise power increases by a factor
\begin{equation}
\langle w^2\rangle=\frac{3}{2}\,.
\end{equation}

Next we translate this result to the appropriate normalization for
our Fourier transform. If we have $N_{\rm bins}$ bins of equal width
$\Delta=\tau/N_{\rm bins}$, and the signal rate $R(t)$, the Fourier
transform of this rate sampled at times $t_j=j\,\Delta$ with
$j=0,\ldots, N_{\rm bins}-1$ is
\begin{equation}
h(f_k) = \Delta \sum_{j=0}^{N_{\rm bins}-1} R(t_j)\,\E^{\I 2 \pi t_j k \delta f}\,.
\end{equation}
The frequencies are $f_k=k/\tau=k\delta f$ with $k=0,\ldots,N_f$ and
$N_f=f_{\rm max}/\delta f$. Here $f_{\rm max}=1/2\Delta$ is the
Nyquist frequency. Since $\delta f=1/\tau$ we have $N_f=N_{\rm
bins}/2$.

In practice our data are provided for a duration $\tau=400$~ms so
that $\delta f=2.5$~Hz. We use 1~ms sampling and thus consider
$\Delta=1$~ms bins, providing a Nyquist frequency of $f_{\rm
max}=500$~Hz as an upper cutoff. For $f=0$ and $f_{\rm max}$ the
spectral power is respectively defined as
\begin{equation}
P(0)=\frac{|h(0)|^2}{N_{\rm bins}^2}
\quad\hbox{and}\quad
P(f_{\rm max})=\frac{|h(f_{\rm max})|^2}{N_{\rm bins}^2}\,.
\end{equation}
For all other frequencies we define
\begin{equation}
P(f_k)=\frac{|h(f_k)|^2+|h(-f_k)|^2}{N_{\rm bins}^2}
=2\,\frac{|h(f_k)|^2}{N_{\rm bins}^2}\,.
\end{equation}
The second equality applies because the transformed function is real
and therefore $|h(-f)|^2=|h(f)|^2$.

Using a signal duration $\tau=400$~ms and
Eq.~(\ref{eq:backgroundrate}) for the background rate we find:
$N_{\rm bkgd}=5.4\times10^{5}$ and $h(0)=N_{\rm bkgd}$. The ratio
$|h(f_k)|^2/|h(0)|^2$ was earlier found to be $1/N_{\rm bkgd}$ times
a factor 3/2 if we use the Hann window. Our definition of power for
binned data involves a factor $2/N_{\rm bins}^2$. Therefore, we find
for the IceCube shot noise, relevant for a signal duration of 400~ms
and a Hann window,
\begin{equation}\label{eq:shotnoise}
P_{\rm shot}=\frac{3}{2}\,\frac{N^2_{\rm bkgd}}{N_{\rm bkgd}}\,
\frac{2}{N_{\rm bins}^2}=
\frac{3N_{\rm bkgd}}{N_{\rm bins}^2}=
10.08\,.
\end{equation}
We have confirmed this result with a few numerical Monte Carlo
realizations.

If we use a subset of the full data, i.e.\ a shorter signal duration
$\tau$ with correspondingly fewer bins, the frequency spacing is
increased, but the power at a given frequency remains the same
except for detailed changes implied by the reduced data. Both
$N_{\rm bkgd}$ and $N_{\rm bins}$ get reduced linearly with $\tau$
and therefore $P_{\rm shot}\propto\tau^{-1}$. The signal-to-noise of
spectral power in a stationary signal increases linearly with
integration time.


\end{document}